\documentclass[11pt]{amsart}
\usepackage[centering]{geometry}                
\usepackage{graphicx}
\usepackage{amssymb}
\usepackage{epstopdf}
\usepackage{pdfpages}

\def\expect{E}

\author{Micha{\l} Komorowski$^{1,*}$, Maria J. Costa$^{2}$,
David A. Rand$^{2}$,
Michael P.H. Stumpf$^{1,*}$
}

\title{Sensitivity, robustness and identifiability in stochastic chemical kinetics models}

\date{\today}                                           

\begin{document}
\pagestyle{plain}
\maketitle
\hspace{10mm}
\begin{minipage}{\textwidth}
\noindent 1.Division of Molecular Biosciences, Imperial College London, UK\\
\noindent 2. Systems Biology Centre,  University of Warwick, UK\\
\ \\{\small
\noindent $*$ Correspondence: {\it M.Komorowski@imperial.ac.uk, M.Stumpf@imperial.ac.uk }   }
\end{minipage}

\begin{abstract}
We present a novel and simple method to numerically calculate Fisher Information Matrices for stochastic chemical kinetics models. The linear noise 
approximation is used to derive model equations and a likelihood function which leads to an efficient computational algorithm. Our approach reduces the problem of calculating the Fisher Information Matrix to solving a set of ordinary differential equations. {This is the first method to compute Fisher Information for stochastic chemical kinetics models without the need for Monte Carlo simulations.} This methodology is then used to study sensitivity, robustness and parameter identifiability in stochastic chemical kinetics models. We show that significant differences exist between stochastic and deterministic models as well as between stochastic models with time-series and time-point measurements.  We demonstrate that these discrepancies arise from the variability in molecule numbers, correlations between species, and temporal correlations and show how this approach can be used in the analysis and design of experiments probing stochastic processes at the cellular level. The algorithm has been implemented as a Matlab package and is available from the authors upon request.
\end{abstract}


\  \\

Understanding the design principles underlying complex biochemical networks cannot be grasped by intuition alone \cite{csete2002reverse}. Their complexity implies the need to build mathematical models and tools for their analysis. {One of the powerful tools to elucidate such systems' performances} is  
sensitivity analysis \cite{varma1999parametric}.  Large sensitivity to a parameter suggests that the system's output can change substantially  with small variation in a parameter. Similarly large changes in an insensitive  parameter will have little effect on the behaviour. 
Traditionally, the concept of sensitivity has been applied to continuous deterministic systems described by differential equations in order to  identify which parameters a given output of the system is most sensitive to; here sensitivities are computed via the integration of the linearisation of the model parameters  \cite{varma1999parametric}.\\
In modelling biological processes, however, recent years have have witnessed rapidly increasing interest in stochastic models \cite{maheshri2007living}, as experimental and theoretical investigations have demonstrated the relevance of stochastic effects in chemical networks \cite{mcadams1997smg, elowitz2002sge}. Although  stochastic models of biological processes are now routinely being applied to study biochemical phenomena ranging from metabolic networks to signal transduction pathways \cite{wilkinson2009smq}, tools for their analysis are in their infancy compared to the deterministic framework. In particular, sensitivity analysis in a stochastic setting is usually, if at all, 
performed { by analysis of a system's mean behaviour } or using computationally intensive Monte Carlo simulations to approximate finite differences of a system's output or the Fisher information matrix with associated sensitivity measures \cite{gunawan2005sensitivity, rathinam2010efficient}. {The Fisher information has a prominent role in statistics and information theory: it is defined as the variance of the score and therefore allows us to measure how reliably inferences are. Geometrically, it corresponds to the curvature around the maximum value of the log-likelihood.}\\
The interest in characterising the parametric sensitivity of the dynamics of biochemical network models has two important reasons.
First, sensitivity is instrumental for deducing system properties, such as robustness (understood as stability of behaviour under simultaneous changes in model parameters) \cite{Rand2007}. The concept of robustness is of significance, in turn, as it is related to many biological phenomena such as canalisation, homeostasis, stability, redundancy, and plasticity \cite{daniels2008sloppiness}.  Robustness is also relevant for characterising the dependence between parameter values and system behaviour. For instance, it has recently been reported that a large fraction of the parameters characterising  a dynamical system are insensitive  and can be varied over orders of magnitude without significant effect on system dynamics \cite{brown2003sma, Rand2006,Erguler2011}. \\
Second,  methods for optimal experimental design use sensitivity analysis to define the conditions under which an experiment is to be conducted in order to maximise the information content of the data \cite{emery1998optimal}.  Similarly, identifiability analysis uses the concept of sensitivity  to determine {\it a priori} whether certain parameters can be estimated from experimental data of a given type \cite{rothenberg1971identification}.\\
We use the linear noise approximation (LNA) as a continuous approximation to Markov jump processes defined by the Chemical Master Equation (CME). This approximation has  previously been used successfully for modelling as well as for inference \cite{JohanElf11012003, ja_LNA, ruttor2009approximate}.  Applying the LNA allows us to represent the Fisher Information matrix (FIM) as a solution of a set of ordinary differential equations (ODEs). We use this framework to investigate model robustness,  study the information content of experimental samples and calculate Cram\'er-Rao (CR)  bounds for model parameters. Analysis is performed for time series (TS) and time point (TP) data as well as for a corresponding deterministic (DT) model. Results are compared with each other and  provide novel insights into the consequences of stochasticity in biochemical systems. Two biological examples are used to demonstrate our approach and its usefulness:  a simple model of gene expression and a model of the p53 system.
We show that substantial differences in the structure of FIMs exist between stochastic and deterministic versions of these models. Moreover, discrepancies appear also between stochastic models with different data types (TS, TP),  and these can have significant impact on sensitivity, robustness and parameter identifiability. We  demonstrate that differences arise from general variability in the number of molecules, correlation between them and temporal correlations.
\section{Chemical kinetics models}
We consider a general system of $N$ chemical species inside a fixed volume  and let $x=(x_1,\ldots ,x_N)^T$ denote the
number of molecules. The stoichiometric matrix
${S}=\{s_{ij}\}_{i=1,2\ldots N;\
j=1,2\ldots R}$ describes changes in the
population sizes due to $R$ different
chemical events, where each $s_{ij}$
describes the change in the number of
molecules of type $i$ from $X_i$ to $X_i +
s_{ij}$ caused by an event of type $j$. The
probability that an event of type $j$ occurs
in time interval $[t,t+dt)$ equals
$f_j(\mathbf{x},\Theta,t) dt$.
The functions $f_j(\mathbf{x},\Theta,t)$ are called
 transition rates and $\Theta=(\theta_1,...,\theta_L )$ is a vector of model parameters. This
specification leads to a Poisson birth and death process with transition densities described
by the CME (see Supplementary Information (SI)).
Unfortunately,  the CME is not easy to analyze and hence various  approximations have been developed. As shown in \cite{JohanElf11012003, ja_LNA,  ja_BJ_inf} the linear noise approximation provides a useful and reliable framework for both modelling and statistical inference. { It is valid for systems with large number of reacting molecules and is an analogy of the Central Limit Theorem for Markov jump processes defined by CME \cite{Kurtz_Realation}.   Biochemical reactions are modelled} through a stochastic dynamic model which essentially approximates a Poisson process by an  ODE model with an appropriately defined noise process. Within the LNA a kinetic model is written  as
\begin{eqnarray}\label{decomp_phi_xi}
x(t)&=&\varphi(t)+\xi(t)\\\label{MRE}
\dot{\varphi}&=&S\ F(\varphi, \Theta,t)\\\label{Xi}
d \xi&=&A(\varphi, \Theta,t)\xi + E(\varphi, \Theta,t) dW, 
\end{eqnarray}
where
\begin{eqnarray}\label{F}
F(\varphi, \Theta,t)=(f_1(\varphi, \Theta,t), ..., f_l(\varphi, \Theta,t) )\\ 
\left\{ A(\varphi, \Theta,t) \right\}_{ik}=\sum_{j=1}^R s_{ij}
\frac{\partial f_j}{\partial \phi_k}\\ \label{A}
{E(\varphi, \Theta,t)}=S\sqrt{diag({{F} (\varphi,\Theta,t)})}.
\end{eqnarray}
Equation (\ref{decomp_phi_xi}) divides the system's state into a macroscopic state, $\varphi(t)=(\phi_1(t),...,\phi_N(t))$, and random fluctuations, $\xi(t)$.  The macroscopic state is described by an ODE (\ref{MRE}), the macroscopic rate equation (MRE), which in general needs to be solved numerically.   Stochastic fluctuations $\xi$  are governed by a {Wiener process ($dW$)} driven  linear stochastic differential equation (\ref{Xi}) with an explicit solution readily available (see SI). 
The variance $V(t)$ of the system's state $x$ can be explicitly written in terms of an ODE  
\begin{equation}\label{variance}
\frac{d V(t)}{dt}=A(\varphi, \Theta,t)V(t) +V(t) A(\varphi, \Theta, t)^T +E(\varphi, \Theta,t)E(\varphi, \Theta, t)^T, 
\end{equation}
which is equivalent to the  fluctuation-dissipation theorem. Similarly, temporal covariances are given by
\begin{equation}\label{cov}
\text{cov}(x(s),x(t))=V(s) \Phi(s,t)^T\ \ \text{for} \ \ \ t \geq s, 
\end{equation}
where $\Phi(s,t)$ is
the fundamental matrix of the non-autonomous
system of ODEs
\begin{equation}\label{fundamental}
\frac{d\Phi(s,t)}{dt}=A(\varphi, \Theta,t)\Phi(s,t),\ \ \ \Phi(s,s)=I.
\end{equation}
Equations (\ref{decomp_phi_xi}-\ref{fundamental}) are used to derive the likelihood of experimental data. {To account for different experimental settings we consider three types of data: time series (TS), time-point (TP) and deterministic (DT). For TS  measurements are taken from a single trajectory (following the same cell) and therefore are statistically dependent; in practise TS data are usually obtained using fluorescent microscopy.  TP measurements at each time point are taken from different trajectories (end time points of  trajectories following different cells) and are thus independent. These data reflect experimental setups where the sample is sacrificed and the sequence of measurements is not strictly associated with the same sample path (e.g. flow-cytometry, Q-PCR). DT data are defined as a solution of MRE (\ref{MRE}) with normally distributed measurement error with zero mean and variance $\sigma^2_{\epsilon}$ and  refer to measurements averaged over population of cells.}  \\
Suppose measurements are collected at times $t_1, ..., t_n$. For simplicity we consider the case where at each time point $t_i$ all components of $x_i$ are measured. In the SI we demonstrate that the same analysis can be done for a  model with unobserved variables at no extra cost other than more complex notation.   First let ${\mathbf{x}_{{\mathit Q}}} \equiv ({x}_{t_1},\ \ldots ,
 {x}_{t_n})$ be  an $nN$ column vector that contains all measurements of type $Q$, where ${\mathit Q}\in \{  TP,TS, DT  \}$. 
 It can be shown (see SI) that  
 \begin{equation}\label{MVN_obs}
\mathbf{x}_{{\mathit Q}} \sim \text{MVN}(\mu(\Theta), \Sigma_{{\mathit Q}}(\Theta))
\end{equation}
where MVN denotes the multivariate normal distribution, 
\begin{equation} \label{mean}
\mu(\Theta)=(\tilde{\varphi}(t_1), ..., \tilde{\varphi}(t_n) ), 
\end{equation}
and $\tilde{\varphi}(t)$ is a solution of the MRE (\ref{MRE}) such that $\tilde{\varphi}(0)=\varphi_0$ and $\Sigma_{{\mathit Q}}$ is a $(nN) \times (nN) $ symmetric block matrix  $\Sigma_Q(\Theta)= \left\{ \Sigma_{{\mathit Q}}(\Theta)^{(i,j)} \right\}_{i=1,...,N;j=1,...,N}$  such that
 
  \begin{equation}\label{Sigma_all_T}
\Sigma_{ {\mathit Q} }(\Theta)^{(i,j)}=
\left\{ 
\begin{array}{cc}
 \tilde{V}(t_i)  & \text{for}\ \   i=j  \ \ \  {\mathit Q}\in \{ TS, TP\} \\
 \sigma^2_{\epsilon} I  & \text{for}\ \   i=j  \ \ \  {\mathit Q}\in \{ DT\}\\
 0  & \text{for} \ \ i < j  \ \ \  {\mathit Q}\in \{TP, DT\}\\
 \tilde{V}(t_i) \Phi(t_i, t_{j})^T  & \text{for} \ \ i < j {\mathit Q} \in \{TS\}
 \end{array}\right.
\end{equation}
\noindent and $\tilde{V}(t)$ is a solution of eq. (\ref{variance}) for a given initial condition $\tilde{V}(0)=V_0$. The MVN likelihood is a result of our LNA and is analogous to the Central Limit Theorem for the CME.  It is valid under the assumption of large number of molecules reacting in the system \cite{Kurtz_Realation}.

\section{Fisher information matrix}
To calculate the FIM  \footnote{In the paper we are interested in the expected FI that under standard regularity conditions is equivalent to the expected Hessian of the likelihood. The expected FI is different from observed FI defined as Hessian of the likelihood of given data.} for the model $(1-3)$, first, suppose that a random variable $X$ has an $N$-variate normal distribution with density $\psi$, mean $\mu(\Theta)=( \mu_1(\Theta), ...,  \mu_N(\Theta))^T$ and covariance matrix $\Sigma(\Theta)$. The FIM is then defined  \cite{porat1986computation} as $I(\Theta)$= $\left\{ I(\Theta)_{k,l} \right\}_{k,l=1,...,L}$, where
\begin{equation}\label{FIM_general}
I(\Theta)_{k,l}=E_{\Theta} \left[ \left( \frac{\partial }{\partial \theta_k}  \log(\psi(X,\Theta )  ) \right) \left( \frac{\partial }{\partial  \theta_l }  \log(\psi(X,\Theta )  ) \right) \right].
\end{equation}
Then $I(\Theta)_{i,j}$  can be expressed as  \cite{frieden2004science}
\begin{equation}\label{FIM_ij}
I(\Theta)_{k,l}=\frac{\partial \mu}{\partial  \theta_k}^T \Sigma(\Theta)\frac{\partial \mu}{\partial  \theta_l}  +\frac{1}{2}trace( \Sigma^{-1}\frac{\partial \Sigma}{\partial  \theta_k} \Sigma^{-1} \frac{\partial \Sigma}{\partial  \theta_l}).
\end{equation}
The above formula shows that, in order to calculate FIM for a multivariate normal distribution, it is enough to calculate the covariance matrix $\Sigma(\theta)$, parameter derivatives of mean $\frac{\partial \mu}{\partial  \theta_k}$ and parameter derivatives of the covariance matrix $\frac{\partial \Sigma}{\partial  \theta_k}$. 

In the LNA equations (\ref{mean}) and (\ref{Sigma_all_T}) describe  mean and variance, respectively,  of experimental measurements, $x_Q$.  The mean is given as the solution of an ODE, and the variance is either given as a product of solutions of ODEs (TS), directly as a solution of an ODE (\ref{variance}) (TP),  or is simply constant (DT). 
Hence, in order to calculate the FIM we calculate the derivatives of the solutions of an ODE with respect to the parameters \cite{coddington1972theory}. For illustration, consider an $N$ dimensional ODE
\begin{equation}\label{aux_z}
\dot{z}=v(z,\theta,t),
\end{equation}
where $\theta$ is a scalar parameter. Denote by $\tilde{z}(z_0,\theta,t)$  the solution of equation (\ref{aux_z}) with initial condition $z_0$ and let $\zeta(t,\theta)=\frac{\partial \tilde{z} } {\partial \theta}$. It can be shown that $\zeta$ satisfies  \cite{coddington1972theory}
\begin{equation}
\dot{\zeta}=J(\tilde{z}(t),\theta,t)\zeta + \frac{\partial}{\partial \theta} v(\tilde{z},\theta,t),
\end{equation}
where $J(\tilde{z}(t),\Theta,t)$ is the Jacobian $\frac{\partial}{\partial z} v(\tilde{z},\theta, t)$.
We can thus calculate derivatives  $\frac{\partial \tilde{\varphi } }{\partial  \theta_k}$,  $\frac{\partial \tilde{V}}{\partial  \theta_k}$ and $\frac{\partial \Phi(t_i, t_j)}{\partial  \theta_k}$
that give $\frac{\partial \mu}{\partial  \theta_k}$ and $\frac{\partial \Sigma}{\partial  \theta_k}$ needed to compute FIM for the model (1-3) (see SI).\\
The FIM is of special significance for model analysis as it constitutes a tool for sensitivity analysis, robustness, identifiability and optimal experimental design as we will show below.
\subsection{The FIM and sensitivity}
{
The classical sensitivity coefficient for an observable $Q$ and parameter $\theta$ is
$$S=\frac{\partial Q}{\partial \theta}.$$
The behaviour of a stochastic system is defined  by observables that are drawn from a probability distribution. The FIM is a measure of how this distribution changes in response to infinitesimal changes in parameters. 
Suppose that  $\ell  (\Theta;X)= log(\psi(X,\Theta) )$ and $\ell (\Theta)=-\expect \left[ \ell  (\Theta;X)\right]$.
Than, 
\begin{equation}
I(\Theta)_{k, l} =
-\expect \left[ \frac{\partial^2 \ell  (\Theta;X)}{\partial \theta_k \partial \theta_l} \right],
\end{equation}
i.e. the FIM is the expected Hessian of $\ell  (\Theta,X)$.
Therefore,  if $\Theta^*$ is the maximum likelihood estimate of a parameter
there is a $L\times L$ orthogonal matrix $C$ such
that, in the new parameters $\theta^\prime = C\  (\Theta -\Theta^*) $,
\begin{equation}\label{ell}
\ell (\Theta) \approx \ell (\Theta^*) -\frac{1}{2} \sum_{i=1}^L \lambda_i \left.\theta^\prime_i\right.^2 
\end{equation}
for $\Theta$ near $\Theta^*$. 
From this it follows that the $\lambda_i$ are the eigenvalues of the
FIM and that the matrix $C$ diagonalises it.
If we assume that the $\lambda_i$ are
ordered so that $\lambda_1 \geq \cdots \geq \lambda_L$
then it follows that around the maximum the likelihood is most sensitive when
$\theta^\prime_1$ is varied and least sensitive when $\theta^\prime_L$ is varied,
and $\lambda_i$ is a measure of this.
Since $\theta^\prime_i =\sum_{j=1}^L C_{ij}(\theta_j -\theta^*_j)$ we can regard
${\mathcal S}_{ij}= \lambda^{1/2}_i\ C_{ij}$ as the contribution of the parameter $\theta_j$
to varying $\theta^\prime_i$ and thus
\begin{equation}
{\mathcal S}_j^2 = \sum_{i=1}^L {\mathcal S}_{ij}^2
\end{equation}
can be regarded
as a measure of the sensitivity of
the system to $\theta_j$. It is sometimes appropriate to normalise
this and instead consider
\begin{equation}
{\mathcal T}_j = \frac{{\mathcal S}_j^2}{\sum_{i=1}^L{\mathcal S}_i^2} .
\end{equation}
\subsection{Robustness}
Related to sensitivity,  robustness in systems biology is usually understood as persistence of a system to perturbations to external conditions \cite{felix2006robustness}. Sensitivity considers perturbation in a single parameter whereas robustness takes into account simultaneous changes in all model parameters.
Near to the maximum $\Theta^*$ the regions of high expected log-likelihood
$\ell (\Theta)\geq \ell (\Theta^*)-\varepsilon$ are approximately
the ellipsoids $NS(\Theta^*, \varepsilon)$ given by the equation
\begin{equation} \label{NP}
NS(\Theta^*, \varepsilon) = \left\{\Theta: (\Theta-\Theta^*)^T I(\Theta^*) (\Theta-\Theta^*) < \varepsilon \right\}.
\end{equation} 
The ellipsoids have principal directions given by eigenvectors $C$ and equatorial radii $(\lambda_i)^{-\frac{1}{2}}$. Sets $NS$ are called neutral spaces as they describe regions of parameter space in which a system's behaviour does not undergo significant changes  \cite{daniels2008sloppiness} and arise naturally in the analysis of robustness.
\subsection{Confidence intervals and asymptotics} {The asymptotic normality of maximum likelihood estimators implies that if  $\Theta^*$ is a maximum likelihood estimator then the NS describe confidence ellipsoids for $\Theta$ with  confidence levels corresponding to $ \varepsilon$. The equatorial radii decrease naturally with the square root of the sample size  \cite{degrootprobability}.}

\subsection{Parameter identifiability and optimal experimental design}
The FIM is of special significance for model analysis as it constitutes a classical criterion for parameter identifiability \cite{rothenberg1971identification}. There exist various definitions of parameter identifiability and here we consider local identifiability. The parameter vector $\Theta$ is said to be (locally) identifiable if there exists a neighbourhood of $\Theta$ such that no other vector  $\Theta^*$ in this neighbourhood gives raise to the same density as $\Theta$ \cite{rothenberg1971identification}. Formula (\ref{ell}) implies that  $\Theta$ is (structurally) identifiable if and only if FIM has a full rank \cite{rothenberg1971identification}. Therefore the number of non-zero eigenvalues of FIM is equal to the number of identifiable parameters, or more precisely, to the number of identifiable linear combinations of parameters. \\
The FIM is also a key tool to construct experiments in such a way that the parameters can be estimated  from the resulting experimental data with the highest possible statistical quality. The theory of optimal experimental design uses various criteria to asses information content of experimental sampling methods; among the most popular are the concepts of D-optimality that maximises the determinant of FIM, and A-optimality that minimise the trace of the inverse of FIM \cite{emery1998optimal}. 
Diagonal elements of the inverse of FIM constitute a lower-bound for the variance of any unbiased estimator of elements of $\Theta$; this is known as the Cram{e}r-Rao inequality (see SI). {Finally, it is important to keep in mind that some parameters may be structurally identifiable, but not be identifiable in practice due to noise; these would correspond to small but non-zero eigenvalues of the FIM. Maximizing the number of eigen-values above some threshold which reflects experimental resolution, may therefore be a further criterion to optimize experimental design. But all of these criteria revolve around being able to evaluate the FIM.
}
}
\section{Results}
In order to demonstrate the applicability of the presented methodology for calculation of FIMs for stochastic  models we consider two examples:  a simple model of  single gene expression, and a model of the p53 system. 
The simplicity of the first model allows us to explain how the differences between deterministic and stochastic versions of the model as well as TS and TP data arise.  In the case of the p53 system model  the informational content, as well as sensitivities and neutral spaces are compared between TS, TP and DT data.
\subsection{Single gene expression model}
\begin{figure}[!h]
\begin{center}
  \includegraphics[scale=0.35]{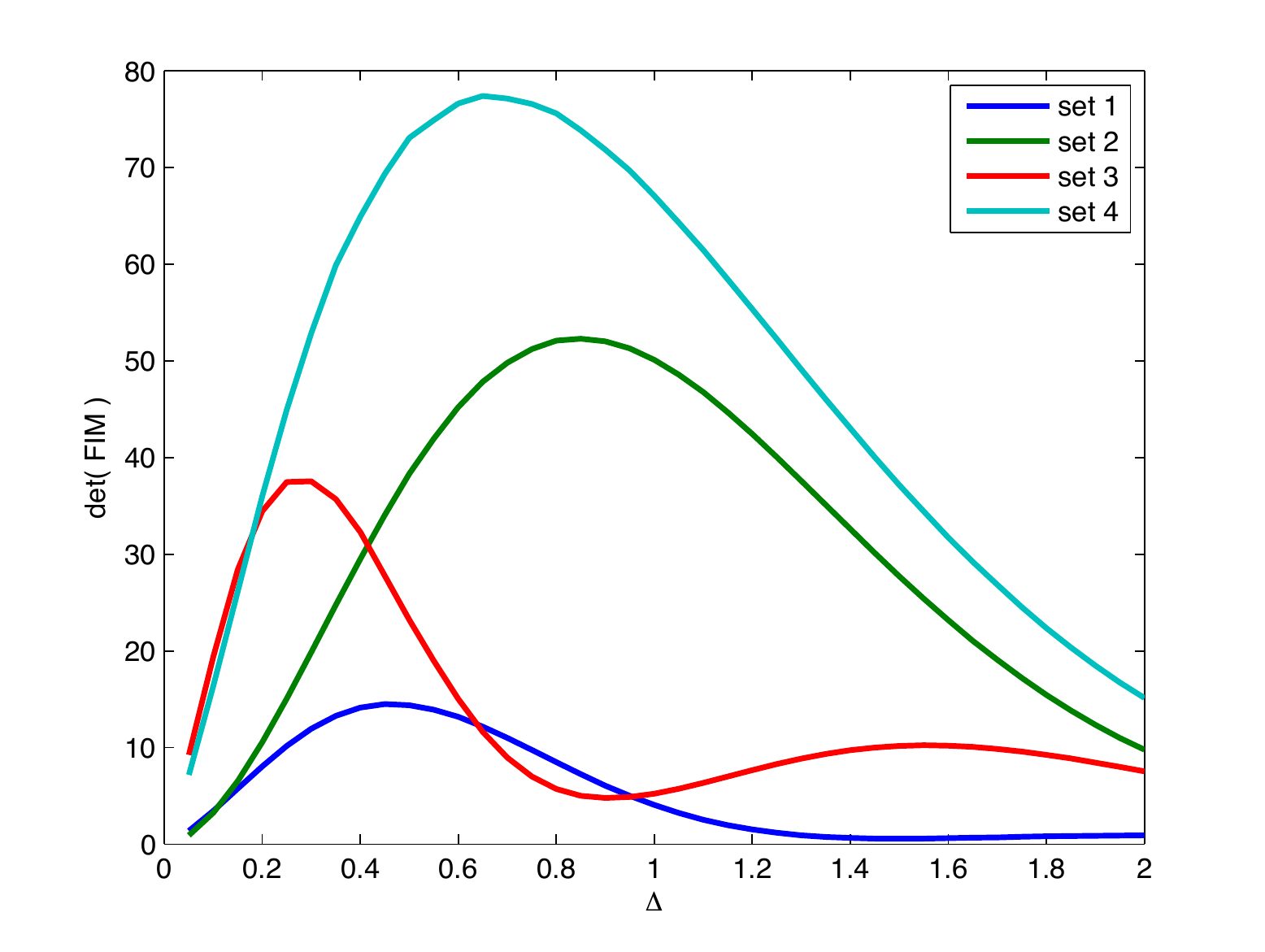}
\end{center}
\caption{\label{FIGURE_gene_det_FIM} Determinant of FIM plotted against sampling frequency $\Delta$ (in hours). We used logarithms of four parameter sets (see {Table 1} SI). Sets 1 and 3 correspond to slow protein degradation $(\gamma_p=0.7)$; and Sets 2 and 4 describe fast protein degradation $(\gamma_p=1.2)$.  We assumed that 50 measurements $(n=50)$ of protein levels were taken from the stationary state. Observed maximum in information content results from the balance between independence and correlation of measurements.}
\end{figure}

Although gene expression involves numerous biochemical reactions, the currently accepted consensus is to model it in terms of only three biochemical species (DNA, mRNA, and protein) and four reaction channels (transcription, mRNA degradation, translation, and protein degradation) (e.g. \cite{MukundThattai07032001,  porat1986computation}). Such a simple model has been used successfully in a variety of applications and can generate data with the same statistical behaviour as more complicated models \cite{dong2006srs, iafolla2006ebp}. We assume that the process begins with the production of mRNA molecules $(r)$ at rate $k_r$. Each mRNA molecule may be independently translated into protein molecules $(p)$ at rate $k_p$. Both mRNA and protein molecules are degraded at rates $\gamma_r$ and $\gamma_p$, respectively. Therefore, we have the state vector  $x=(r, p)$,  
and reaction rates corresponding to transcription of mRNA, translation, degradation of mRNA, and degradation of protein.
\begin{equation}
F(x)=(k_r,k_p r, \gamma_r r, \gamma_p p ).
\end{equation}
\noindent {\it{Identifiability study.}}
In a typical experiment 
only protein levels are measured \cite{ja_LNA,  Oude_Clock}. It is not entirely clear {\it a priori}  what parameters of gene expression can be inferred; it is also not obvious if and how the answer depends on the nature of the data (i.e. TS, TP or DT).  We address these questions below.\\
\noindent We assumed that the system has reached the unique steady state defined by the model and that
only protein level is measured either as TS
\begin{equation}\label{data_gene_TS}
\mathbf{y}_{TS}=({p}_{t_1},...,{p}_{t_n}) 
\end{equation}
or as TP 
\begin{equation}\label{data_gene_TP}
\mathbf{y}_{TP}=({p}_{t_1}^{(1)},...,{p}_{t_n}^{(n)}),
\end{equation}
where the upper indices for TP measurements denote the number of trajectories from which the measurement have been taken to emphasise independence of measurements.  
Results of the analysis are presented in { Table 2 SI}. For TS data we have four identifiable parameters whereas time-point measurements provide enough information to estimate only two parameters. To some extent this makes intuitive sense:  TS data contain information about mean, variance and autocorrelation functions, which can be very sensitive to changes in degradation rates; TP measurements reflect only information about mean and variance of protein levels therefore only two parameters are identifiable. On the other hand TP measurements provide independent samples which is reflected in lower Cram{e}r-Rao bounds.  {Table 2} SI contains also a comparison with the corresponding deterministic model. As one might expect in the deterministic model only one parameter is identifiable as the mean is the only quantity that is described by the deterministic model, and parameter estimates are informed neither by variability nor by autocorrelation.\\
\noindent {\it{Perturbation experiment.}} In order to demonstrate that identifiability is not a model specific but rather an experiment specific feature we performed a  similar analysis as above for the same model with the same parameters but with the 5 fold increased initial mean and 25 fold increased initial variance. Results are presented in {Table 3} SI.  Some of the conclusions that can be made are hard to predict without calculating the FIM. The amount of information in TS data is now much larger than in TP data (higher determinant) and also CR bounds are now much lower for TP than for TS data. CR bounds for TS and TP are substantially lower than for the steady state data  (except $k_r$).  Interestingly,  all four parameters can be inferred from TS and TP data, but not in the deterministic scenario. For steady state data all parameters could only be  inferred from TS data ({Table 3} SI ).\\
\noindent {\it{Maximising the information content of experimental data.}} The amount of information in a sample does not depend solely on the type of data (TS, TP), but also on other factors that can be controlled in an experiment. One easily controllable quantity is the sampling frequency $\Delta$. {We consider here only equidistant sampling and keep number of measurements constant. Therefore we define $\Delta$ as time between subsequent observations $\Delta=t_{i+1}-t_i$.} To show how sampling frequency influences  informational content of a sample for the model of gene expression we used four parameter sets ({Table 1} SI) and assumed that the data have the form (\ref{data_gene_TS}).  The amount of information in a sample was understood as the determinant of the FIM, equivalent to the product of the eigenvalues of the FIM. Results in Figure \ref{FIGURE_gene_det_FIM}  demonstrate that our method can be used to determine optimal sampling frequency, given that at least  some rough estimates of model parameters are known.   { It is worth noting that equidistant sampling is not always the best option and more complex strategies have been proposed in experimental design literature.}

\begin{figure}[!h]
\begin{center}
  \includegraphics[angle=0, height=50mm, width=80mm ]{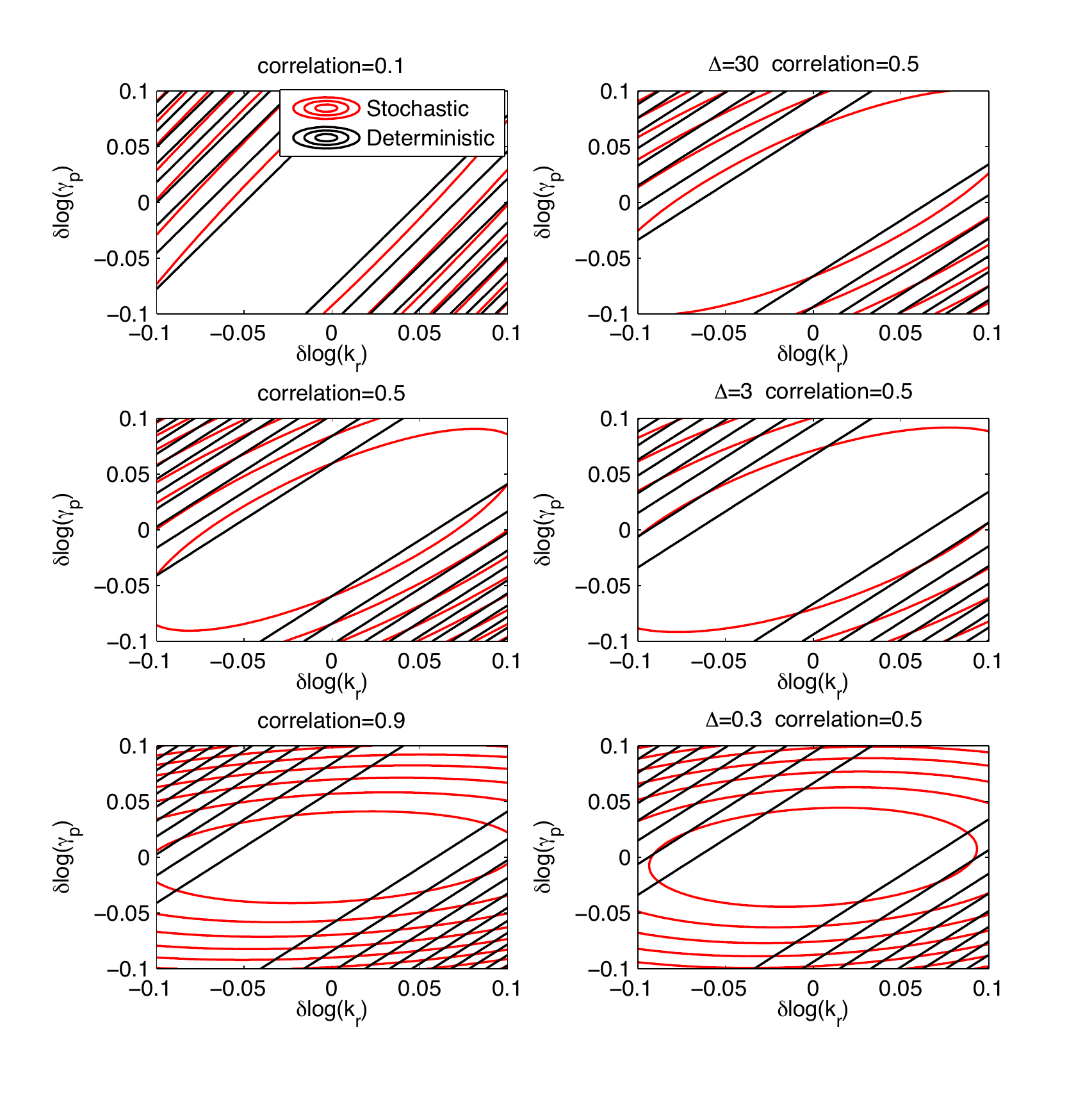}
 \end{center}
\caption{Neutral spaces for TS and DT versions of the model of single gene expression for logs of parameters $k_r$ and $\gamma_p$.
{\bf Left: }Differences resulting from  RNA, protein correlation: $\rho_{rp}=0.1$ (top)   $\rho_{rp}=0.5$  (middle), $\rho_{rp}=0.9$ (bottom). 
Correlation $0.5$ corresponds to parameter set 3 from {Table 1} SI and was varied by equal-scaling of parameters $k_p, \gamma_p$.   {\bf Right:} Differences resulting from temporal correlations. We assumed $n=50$  and tuned correlation between observation by changing sampling frequency  $\Delta=0.3$h (left) $\Delta=3$h (middle) $\Delta=30$h (right).   Set 3 of parameters was used  ({Table 1} SI). 
\label{gene_slop} }
\end{figure}

\noindent  {\it{Differences in sensitivity and robustness analysis in TS, TP and DT versions of the model.}}  TS, TP and DT versions of the model differ when one considers information content of samples, and such discrepancies exist also when sensitivity and robustness are studied.  First, deterministic models completely neglect variability in molecular species. Variability, however, is a function of parameters, and like the mean, is sensitive to them.  Second, deterministic models do not include correlations between molecular species. Third, temporal correlations are neglected in TP and DT models.\\
To understand these effects we first analyse the analytical form of means, variances and correlations for this model (see SI). 
We start with the effect of incorporating variability. Suppose we consider a change in parameters, e.g. $k_p, \gamma_p$ by a factor $\delta$ $(k_p, \gamma_p) \rightarrow (k_p+\delta k_p, \gamma_p +\delta \gamma_p) $. The means of RNA and protein concentrations are not affected by this perturbation, whereas the protein variance does change (see formulae {(33-37)} in SI). This  result is related to the number of non-zero eigenvalues of the FIM. The FIM for the stationary distribution of this model with respect to parameters $k_p, \gamma_p$ has only one positive eigenvalue for the deterministic model and two positive eigenvalues for the stochastic model.\\
To study the effect of correlation between RNA and protein levels $\rho_{rp}$ we first note that formulae { (33 - 37) in SI} demonstrate that at constant mean, correlation increases with $\gamma_p$ when accompanied by a compensating increase in $k_p$. Figure \ref{gene_slop} (left column)
 presents neutral spaces (\ref{NP}) for parameter pairs for different values of correlation, $\rho_{rp}$. The  differences between DT and TS are enhanced by the correlation.\\
Similar analysis reveals that taking account of the temporal correlations also changes the way the model responds to parameter perturbations. 
Figure \ref{gene_slop} (right column) shows neutral spaces for three different sampling frequencies and indicates that the differences between stochastic and deterministic models decrease with $\Delta$.
 
\subsection{Model of p53 system}
\begin{figure}[!h]
\begin{center}
 \includegraphics[width=80mm, height=40mm]{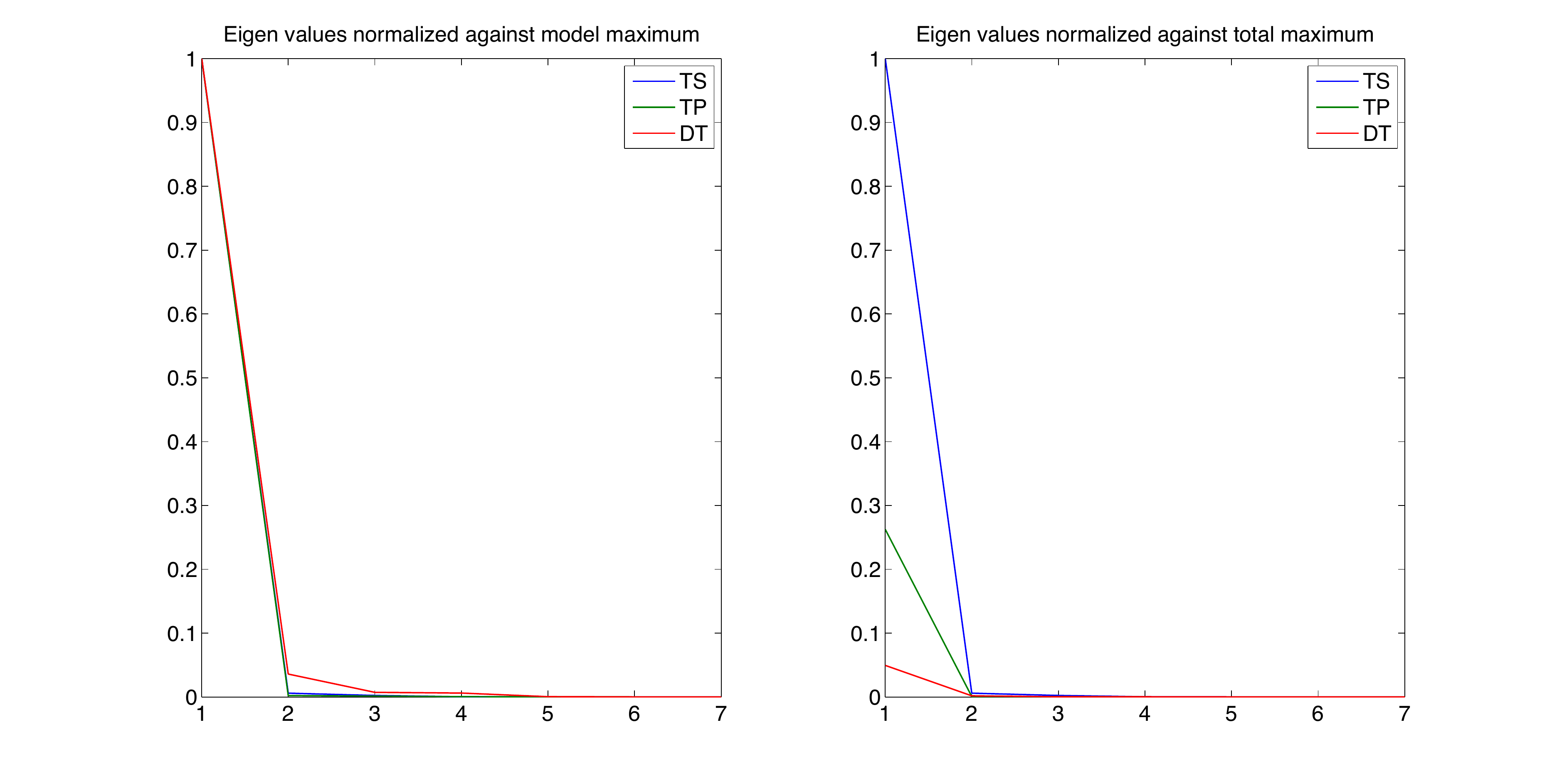}
\caption{ Eigenvalues of FIM for p53 model for three data types: time series (blue), time-points (green) and deterministic model (red). Eigen values were normalised against maximal eigenvalue for each data type (top) and against maximal eigenvalue among all three types (bottom). FIM was calculated for logs of parameters from { Table 4 SI }. \label{p53ev} }
\end{center}
\end{figure}

\begin{figure}[!h]
\begin{center}
\includegraphics[angle=0, scale=0.2] {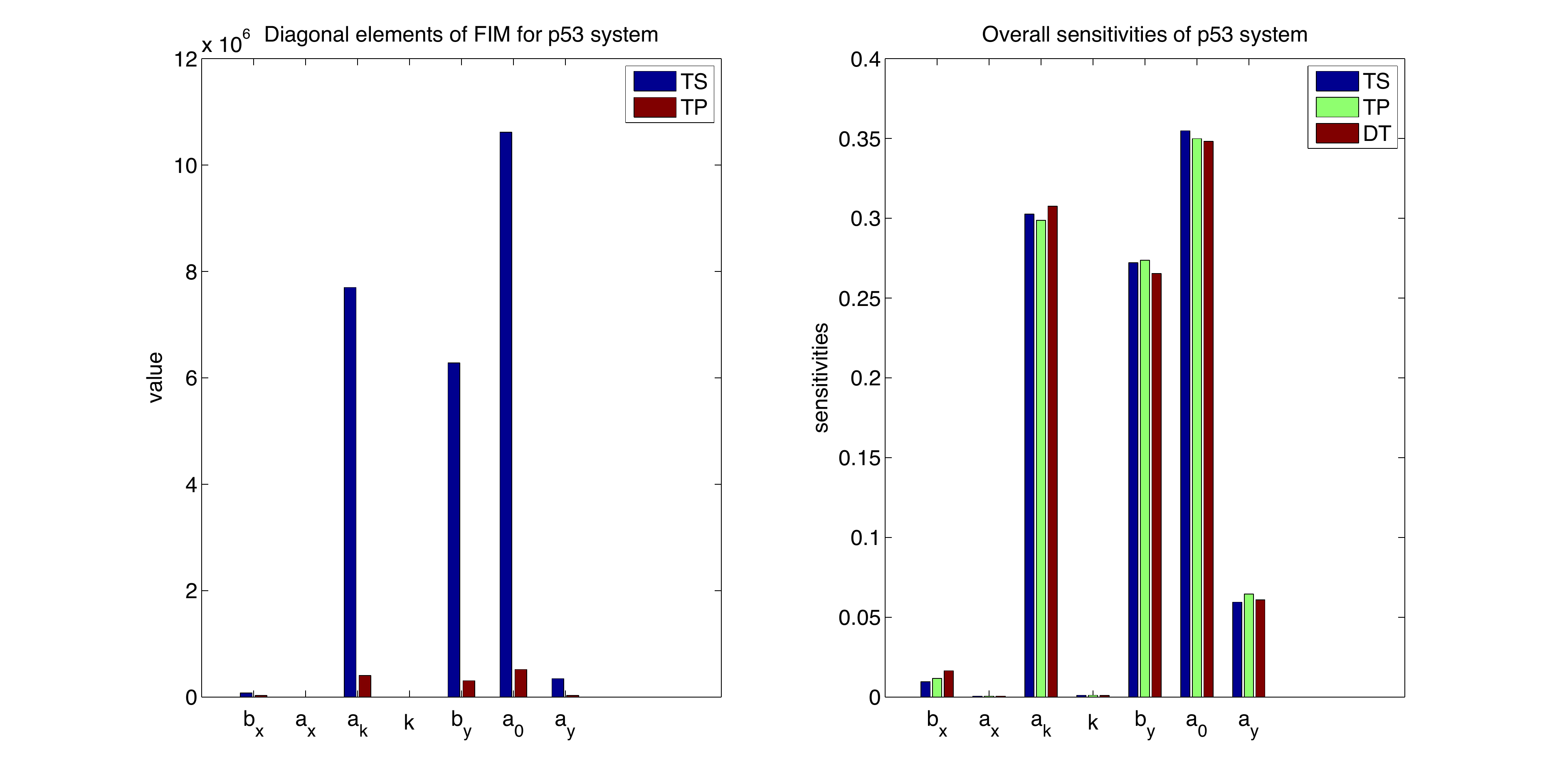}
\end{center}
\caption{ {\bf Left:} Diagonal elements of FIM for TS and TP versions of p53 model.  Values of FIM for DT verison are not presented as they can not be compared with those for stochastic models.  {\bf Right: } Sensitivity coefficients $\mathcal{T}_{i}$ for TS, TP, DT version of p53 model. FIMs were calculated for parameters presented in Table 4 SI. \label{p53_diag_sens} }
\end{figure}

The model of single gene expression is a linear model with only four parameters and a simple stationary state and illustrates how the methodology can be used to provide relevant conclusions and investigate discrepancies between sensitivities of TS, TP and DT models. Our methodology, however, can also be used to study more complex models, and here we have chosen the p53 signalling system, which incorporates a feedback loop between the tumour suppressor p53 and the oncogene Mdm2, and is involved in regulation of  cell cycle and response to DNA damage.\\
We use the model introduced in \cite{geva2006oscillations} that reduces the system to three molecular species, p53, mdm2 precursor and  mdm2, denoted here by $p, y_0$  and $y$, respectively.
The state of the system is therefore given by
$x=(p,y_0,y)$, and  
{  
the seven reactions present in the model are described by the stoichiometry matrix 
\begin{equation}
S=
\left(
\begin{array}{cccccc}
1& -1& -1 &0& 0& 0\\
0& 0 &0 &1 &-1& 0\\
0& 0 &0& 0& 1& -1\\
\end{array}
\right)
\end{equation}
and occur at rates 
$$F(x)=(b_x,a_xp, a_{k}y\frac{p}{p+k},b_yp, a_0 y_0, a_yy),$$
so that the vector of model parameters can be written as
$$\Theta=(b_x,a_x, a_{k}, k, a_y, a_0, a_y).$$
The above specification allow us to formulate a macroscopic rate equation model
\begin{eqnarray}
\dot{\phi}_p&=& b_x-a_x \phi_p -a_{k}\phi_y\frac{\phi_p}{\phi_p+k}\\
\dot{\phi}_{y_0}&=&b_y \phi_p- a_0 \phi_{y_0} \\
\dot{\phi}_y&=& a_0 \phi_{y_0} - a_y \phi_{y}.
\end{eqnarray}}
\noindent {\it{ Informational content of TS and TP data for the p53 system.}}
{In the case of the single gene expression model we have argued that TS data are more informative due to accounting for temporal correlations. On the other hand TP measurements provide statistically independent samples, which should increase informational content of the data. Therefore it is not entirely clear what data type is better for a particular parameter. If, for instance, a parameter is entirely  informed by a system's mean behaviour than TP data will be more informative because TP data provide statistically independent samples about the mean. Whereas if a parameter is also informed by temporal correlations, then TS data will turn out to be more informative.  It is difficult to predict {\em a priori} which effect will be dominating. Therefore calculation of FIM and comparison of their eigenvalues and diagonal elements is necessary. }Eigenvalues and diagonal elements of FIMs calculated for parameters presented in {Table 4} SI are plotted in Figure \ref{p53ev} and Figure \ref{p53_diag_sens}, respectively. Eigenvalues of the FIM for TS data are larger than for TP data. Similarly, diagonal elements for all parameters are larger for TP than for TS data for most parameters difference is substantial. This indicates that temporal correlation is a sensitive feature of this system and provides significant information  about model parameters. The lower information content of the TP data can, however, be compensated for by increasing the number of independent measurements, which is easily achievable in current experimental settings (see Figure 7 SI). For deterministic models  the absolute value of elements of FIM depends on measurement error variance and therefore FIMs of TS and TP data can not be directly compared with  the DT model.\\
\noindent {\it{ Sensitivity.}}
The sensitivity coefficients $\mathcal{T}_i$ for TS, TP and DT data are presented in Figure \ref{p53_diag_sens}. Despite differences outlined previously, here sensitivity coefficients are quite similar for all three types suggesting that the hierarchy of  sensitive parameters is to a considerable degree  independent on the type of data. The differences exist, however, in contributions $C_{ij}^2$ (see {Figure 6} in SI), suggesting discrepancies in neutral spaces and robustness analysis that we present below.\\
\noindent {\it{ Neutral spaces.}} 
 Comparison of the neutral spaces (\ref{NP}) for each pair of data types and for each pair of the parameters are given in {Figures 3-5} SI. The conclusion we can draw from these figures is that NSs for TS, TP and DT model exhibit substantial differences; these differences, however, are limited to certain parameter pairs. Differences between NPs of TS and DT models are exhibited in pairs involving parameters  $b_x$, $a_y$; between TS and TP in pairs involving $b_x$; and between TP and DT also pairs involving $b_x$.\\
This suggests that parameter $b_x$ is responsible either for the variability in molecular numbers or the correlation between species, as these are  responsible for differences between TP and DT models. Similarly the lack of differences in pairs involving $a_y$ in comparisons of TP and DT, and their presence in comparison of TP and TS indicates that parameter $a_y$ is responsible for regulating the temporal correlations. This analysis agrees with what one might intuitively predict. Parameter $b_x$ describes the production rate, and therefore  the mean expression level of p53, and  also the variability of all components of the system. It is difficult, however, to say how this parameter influences  correlations between species. Parameter $a_y$, on the other hand, is the degradation rate of mdm2 and therefore clearly determines the temporal correlation of not only mdm2 but also of p53, because mdm2 regulates the degradation rate of p53.  While  heuristic, our analysis of the neutral spaces nevertheless clearly demonstrates the differences between the three types of models and creates a theoretical framework for investigating the role of parameters in the stochastic chemical kinetics systems { and without the need to perform Monte Carlo sampling or other computationally expensive schemes}.

\section{Discussion}
The aim of this paper was to introduce a novel theoretical framework that allows us to gain insights into sensitivity and robustness of stochastic reaction systems through analysis of  the FIM. We have used the linear noise approximation \cite{vanKapmen, JohanElf11012003, ja_LNA} to model means, variances and correlations in terms of appropriate ODEs. Differentiating  the solution of these ODEs with respect to parameters \cite{coddington1972theory} allowed us to numerically calculate derivatives of means, variances and correlations,  which combined with the normal distribution of model variables implied by the LNA gave us the representation of the FIM in terms of solutions of ODEs. To our  knowledge this is  the first method to compute FIM for stochastic chemical kinetics models without the need for Monte Carlo simulations.\\
Given the role of the FIM in model analysis and increasing interest in stochastic models of biochemical reactions, our approach is widely applicable. It is primarily aimed at optimising or guiding experimental design, and here we have shown how it can be used to test parameter identifiability for different data types, determine optimal sampling frequencies, examine information content of experimental samples and calculate Cram{e}r-Rao bounds for kinetic parameter estimates.  Its applicability, however, extends much further: it can provide a rationale as to which variables should be measured experimentally, or what perturbation should be applied to a system in order to obtain relevant information about parameters of interest. Similar strategies can also be employed in order to optimise model selection procedures. As demonstrated here stochastic data incorporating information about noise structure are more informative and therefore experimental optimisation for stochastic models models may be advantageous  over similar methods for deterministic models.\\
A second topical application area is the study of robustness of stochastic systems. Interest in robustnesses results from the observation that biochemical systems exhibit surprising stability in function under various environmental conditions. For deterministic models this phenomenon has been partly explained by the existence of regions in parameter space (neutral spaces) \cite{daniels2008sloppiness}, in which perturbations to parameters do not result in significant changes in system output.  We have demonstrated that even a very simple stochastic linear model of gene expression exhibits substantial differences when its neutral spaces are compared with the deterministic counterpart. Therefore a stochastic system may respond differently to changes in external conditions than  the corresponding deterministic model. Our study presents examples of changes in parameters that do not affect behaviour of a deterministic systems but may substantially change a probability distribution that defines the behaviour of the corresponding stochastic system. Thus for systems in which stochasticity plays an important role random effects can not be neglected when considering issues related to robustness.\\
\ \\
{\bf Acknowledgments}
MK and MPHS acknowledge support from the BBSRC (BB/G020434/1).
 DAR holds an EPSRC Senior Fellowship (GR/S29256/01), and his work and that of MC were funded by a BBSRC/EPSRC SABR grant (BB/F005261/1, ROBuST project). DAR and MK were also supported by the European Union (BIOSIM Network Contract 005137). MPHS is a Royal Society Wolfson Research Merit Award holder.

\begin{figure}[!h]
\begin{center}
\includegraphics[angle=0, scale=0.43] {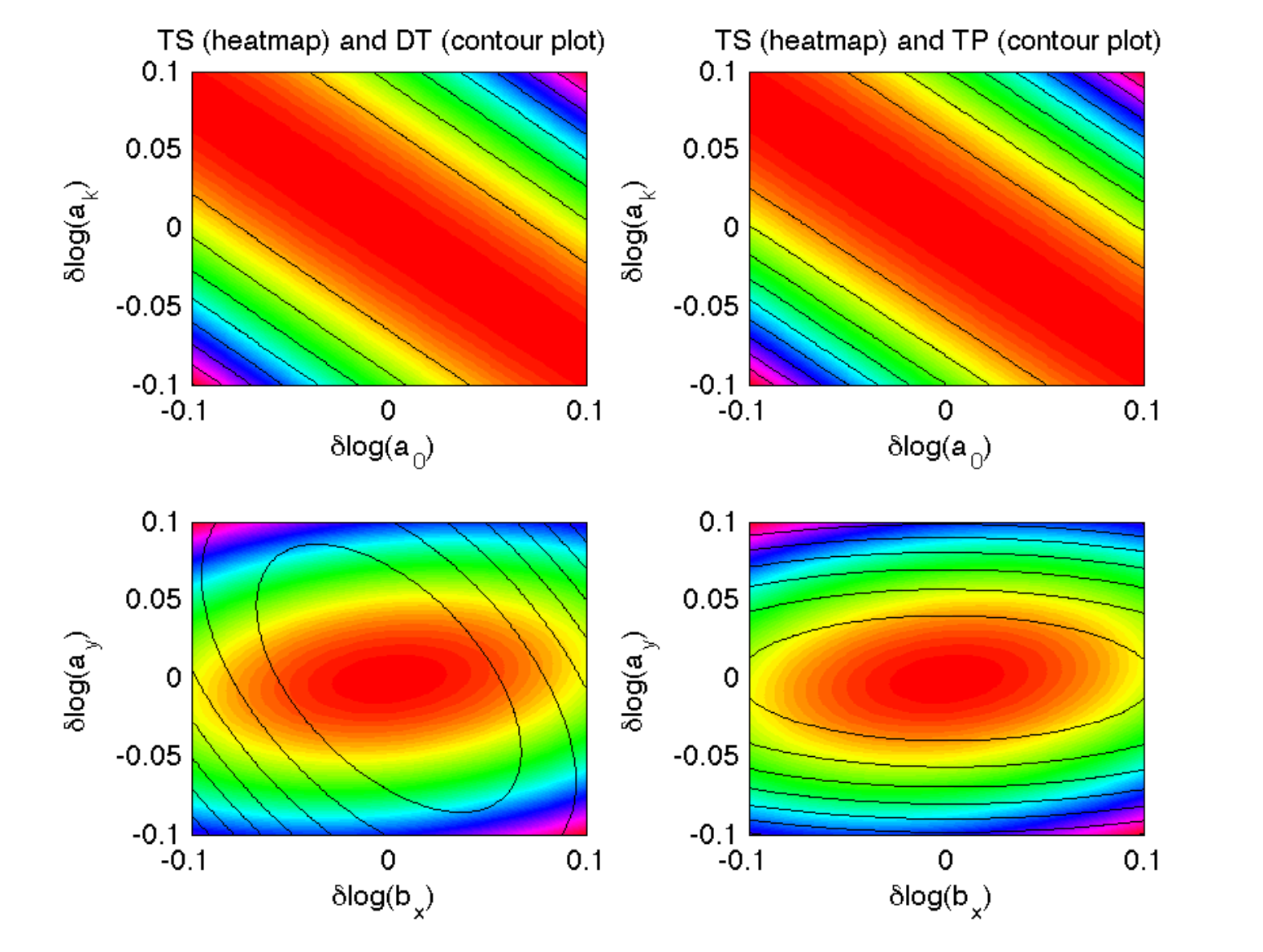}
\end{center}
\caption{ Neutral spaces for TS, TP, and DT versions of p53 model  for logs of two parameter pairs  $(a_0,a_k)$ and $(b_x, a_y)$.
Left column presents differences resulting form general variability, correlations between species and temporal correlation (comparison of TS and TP models). Right column shows differences due to variability and correlation between species (comparison of TS and TP models). Top row is an example of parameters for which differences are negligible,  bottom row presents a parameter pair with substantial differences. FIM was calculated for  30 measurements of all model variables and $\Delta=1$h.
\label{p53_slop_delta}
}
\end{figure}


%
\ \\
\
 
\ \\
\ 

\ \\

\ \\
\
 
\ \\
\ 

\ \\

\ 

\ \\
\ 

\ \\



\section*{Supplementary material}

# Supplementary Information

## Sensitivity, robustness and identifiability in stochastic chemical kinetics models

Michał Komorowski[1], Maria J. Costa[2], David A. Rand[2], Michael Stumpf[1]

1. Centre for Bioinformatics, Imperial College London, UK
2. Systems Biology Centre, and Mathematics Institute, University of Warwick, UK

# 1 Derivation of the model equations

We consider a general system of chemical reactions that consists of $N$ chemical species and interacts in a fixed volume through $R$ reactions. Let $x = (x_1, \ldots, x_N)^T$ be the vector representing the numbers of molecules for the $N$ species and $S = \{s_{ij}\}_{i=1,2\ldots N;\ j=1,2\ldots R}$ be the stoichiometry matrix that describes changes in the population sizes due to each of the reactions, so that occurrence of reaction $j$ results in a change

$$(x_1, \ldots, x_N) \rightarrow (x_1 + s_{1j}, \ldots, x_N + s_{Nj}).$$

We assume that the probability that a reaction of type $j$ occurs in the time interval $[t, t+dt)$ equals $f_j(x, \Theta)dt$, where functions $f_j(x, \Theta)$ are called the reaction rates and $\Theta = (\theta_1, \ldots, \theta_L)$ is the vector of all model parameters. The probability that more than one event will take place in a small time interval is of higher order $(dt^2)$ with respect to the length of the interval and can thus be ignored. Finally, we assume that events taking place in disjoint time intervals are independent, when conditioned on the events in the previous interval. This specification leads to a Poisson birth and death process; the Chemical Master Equation [1, 2] is widely used to describe the temporal evolution of the probability $P(x, t)$ that the system is in the state $x$ at time $t$

$$\frac{dP(x,t)}{dt} = \sum_{j=1}^{R} \left( P(x - s_{\cdot j}, t) f_j(x - s_{\cdot j}, \Theta, t) - P(x,t) f_j(x, \Theta, t) \right). \tag{1}$$

Under the assumption that molecular species are present in sufficiently large copy numbers the model defined above is well described by the following system of equations [1, 3, 4]

$$
\begin{align}
x(t) &= \varphi(t) + \xi(t) \tag{2}\\
\dot{\varphi} &= S\,F(\varphi, \Theta, t) \tag{3}\\
d\xi &= A(\varphi, \Theta, t)\xi + E(\varphi, \Theta, t)dW, \tag{4}
\end{align}
$$

where

$$
F(\varphi, \Theta, t) = (f_1(\varphi, \Theta, t), ..., f_R(\varphi, \Theta, t)) \tag{5}
$$

$$
\{A(\varphi, \Theta, t)\}_{ik} = \sum_{j=1}^{R} S_{ij}\frac{\partial f_j}{\partial \phi_k} \tag{6}
$$

$$
E(\varphi, \Theta, t) = S\sqrt{diag(F(\varphi, \Theta, t))}. \tag{7}
$$

Equation (3) is an ordinary differential equation that in general does not have an explicit solution but can be solved numerically, whereas equation (4) is a linear stochastic differential equation that has a solution of the form [5]:

$$
\xi(t) = \Phi(t_0, t)\xi_{t_0} + \int_{t_0}^{t} \Phi(s, t)E(\varphi, \Theta, s)dW(s), \tag{8}
$$

where the integral is in the Itô sense and $\Phi(t_0, s)$ is the fundamental matrix of the non-autonomous system of ODEs

$$
\frac{d\Phi(t_0, s)}{ds} = A(\varphi, \Theta, s)\Phi(t_0, s), \quad \Phi(t_0, t_0) = I. \tag{9}
$$

In order to simplify the further analysis of the system studied, suppose that the initial condition has a multivariate normal distribution (MVN) $x(0) \sim MVN(\varphi(0), V(0))$.

This specification of an initial condition together with equations (2 - 4, 8) implies that $x(t)$ has a multivariate normal distribution [5, 6]

$$
x(t) \sim MVN(\varphi(t), V(t)) \quad t > 0, \tag{10}
$$

where $\varphi(t)$ is a solution of the macroscopic rate eqaution (MRE), Eqn. (3), with initial condition $\varphi(0)$, and $V(t)$ is a variance at time $t$. Direct calculations using equations (2 - 4, 8) show that $V$ satisfies

$$
\frac{dV(t)}{dt} = A(\varphi, \Theta, t)V(t) + V(t)A(\varphi, \Theta, t)^T + E(\varphi, \Theta, t)E(\varphi, \Theta, t)^T, \tag{11}
$$

which is equivalent to the fluctuation dissipation theorem [1]. In the further sections we will also need to specify covariances, $\text{cov}(x(s), x(t))$ $(t > s)$, and therefore we calculate these here. As $\langle x(t) \rangle = \varphi(t)$ we have that $\text{cov}(x(s), x(t)) = \langle \xi(t)\xi(s)^T \rangle$ and therefore equation (8) implies

$$\text{cov}(x(s), x(t)) = V(s)\Phi(s, t)^T. \tag{12}$$

## 2 Derivation of the likelihood function

In the previous section we have explained that $x(t) \sim MVN(\varphi(t), V(t))$. Now we derive the distribution of experimental data. Three different data types are considered: time series, time-point measurements, and deterministic model data.

### 2.1 Time series data

We start with the case where a single trajectory is measured at times $t_1, ..., t_n$. Initially suppose that all molecular species $x_i$ are measured. Later we demonstrate that this assumption is easily relaxed. First let $\mathbf{x}_{TS} \equiv (x_{t_1}, \ldots, x_{t_n})$ be a $nN$ column vector that contains all measurements and $\tilde{\varphi}(\varphi_0, \Theta, t)$ be a solution of equation (3) such that $\tilde{\varphi}(\varphi_0, \Theta, 0) = \varphi_0$, and let $\tilde{V}(V_0, \Theta, t)$ be a solution of equation (11 ) such that $\tilde{V}(V_0, \Theta, 0) = V_0$. In order to find the distribution of vector $\mathbf{x}_{TS}$ we write $x_{t_0} = \varphi(t_0) + \varsigma_{t_0}$, where $\varsigma_{t_0} \sim MVN(0, V_0)$ and using equations (3-4) and (8) we have

$$x_{t_1} = \varphi_{t_1} + \Phi(t_0, t_1)\varsigma_{t_0} + \varsigma_{t_1},$$

where $\varsigma_{t_1} \sim (0, \Xi_1)$ and $\Xi_1 = \int_{t_0}^{t_1} \Phi(s, t_1)^T E(s)^T E(s) \Phi(s, t_1) ds$. Using

$$\Phi(t_{j-1}, t_{j+1}) = \Phi(t_j, t_{j+1})\Phi(t_{j-1}, t_j) \tag{13}$$

we can analogously write $x_{t_i}$ as

$$x_{t_i} = \varphi_{t_i} + \sum_{j=0}^{i} \Phi_{t_j}(t_i - t_j)\varsigma_{t_j}, \tag{14}$$

where $\varsigma_{t_j}$ are independently normally distributed random variables with mean 0 and covariance matrix $\Xi_j = \int_{t_{j-1}}^{t_j} \Phi(s, t_j)^T E(s)^T E(s) \Phi(s, t_j) ds$. This representation demonstrates that $x_{t_i}$ is a linear sum of multivariate normal variables and therefore $\mathbf{x}_{TS}$ has a multivariate normal distribution with mean $\mu(\Theta)$ and covariance matrix $\Sigma_{TS}(\Theta)$

$$\mathbf{x}_{TS} \sim MVN(\mu(\Theta), \Sigma_{TS}(\Theta)) \tag{15}$$

3

where $\mu(\Theta) = (\tilde{\varphi}(t_1), ..., \tilde{\varphi}(t_n))$ and $\Sigma_{TS}(\Theta)$ is a $(n \times N) \times (n \times N)$ block matrix $\Sigma_{TS}(\Theta) = \left\{ \Sigma(\Theta)^{(i,j)} \right\}_{i=1,...,N; j=1,...,N}$ such that diagonal elements contain variances $\Sigma(\Theta)^{(i,i)} = \tilde{V}(t_i)$ and non-diagonal elements $(i \neq j)$ covariances $\Sigma(\Theta)^{(i,j)} = \mathrm{cov}(x(t_i), x(t_j))$. Diagonal elements are given by the solution $\tilde{V}$. From representation (14) we have

$$\Sigma^{i,j+1} = \Sigma^{i,j} \Phi(t_j, t_{j+1})^T, \tag{16}$$

which demonstrates that non-diagonal elements can be easily computed from diagonal elements given by solutions of equation (11).

## 2.2 Time-point measurements

Here we consider the case where in an experiment at each time point $t_1, ..., t_n$ a different trajectory is measured. Therefore, measurements come from the same process $x(t)$ but from its independent realisations. We define the measurement vector as $\mathbf{x}_{TP} \equiv (x_{t_1}^{(1)}, \ldots, x_{t_n}^{(n)})$. Upper indices indicate the number of trajectories from which the measurements were taken in order to emphasis that each measurement is taken from a different trajectory. The distribution of $x_{t_i}^{(i)}$ is given by (10). All measurements are independent so that $\mathrm{cov}(x_{t_i}, x_{t_j}) = 0$ for $i \neq j$, therefore

$$\mathbf{x}_{TP} \sim MVN(\mu(\Theta), \Sigma_{TP}(\Theta)) \tag{17}$$

where $\mu(\Theta) = (\tilde{\varphi}(t_1), ..., \tilde{\varphi}(t_n))$ and $\Sigma_{TP}(\Theta)$ has the same diagonal blocks as $\Sigma_{TS}(\Theta)$ and non-diagaonal blocks are equal to 0

$$\Sigma_{TP}(\Theta)^{(i,j)} = \begin{cases} \tilde{V}(t_i) & \text{for } i = j \\ 0 & \text{for } i \neq j. \end{cases} \tag{18}$$

## 2.3 Deterministic model

In order to study differences between stochastic and deterministic regimes we also consider a deterministic model where the system state is described entirely by the MRE (3). In such a model measurements are usually assumed to have the form

$$x(t_i) = \varphi(t_i) + \epsilon_{t_i},$$

, where $\epsilon_{t_i}$ is a normally and independently distributed measurement error with mean 0 and constant variance $\sigma_\epsilon^2$. We denote the measurements for this model by $\mathbf{x}_{DT} \equiv (x_{t_1}, \ldots, x_{t_n})$. Finding the data distribution for this case is straightforward,

$$\mathbf{x}_{DT} \sim MVN(\mu(\Theta), \Sigma_{DT}), \tag{19}$$

where $\mu(\Theta)$ is as in the previous cases and $\Sigma_{DT}$ is a $N^2 n^2$ diagonal matrix with diagonal elements equal to $\sigma_\epsilon^2$.

## 2.4 Hidden variables

Usually it is not possible to measure all variables present in the system of interest experimentally. Hence we here demonstrate that the distribution of observed components can be directly extracted from the distributions (15, 17, 19). For simplicity we consider the case of time series data only; analysis for the two other data types proceeds analogously. First, we partition the process $x(t)$ into those components $y(t)$ that are observed and those $z(t)$ that are unobserved. Let $\mathbf{y}_{TS} \equiv (\mathbf{y}_{t_1}, \ldots, \mathbf{y}_{t_n})$ and $\mathbf{z}_{TS} \equiv (\mathbf{z}_{t_1}, \ldots, \mathbf{z}_{t_n})$ denote the time series that of $y$ and $z$, respectively, at times $t_1, \ldots t_n$.

The distribution of $\mathbf{y}_{TS}$ is a marginal distribution of $\mathbf{x}_{TS}$; we thus have

$$\mathbf{y}_{TS} \sim MVN(\mu_y(\Theta), \hat{\Sigma}(\Theta)), \tag{20}$$

where $\mu_y(\Theta)$ and $\hat{\Sigma}(\Theta)$ are elements of $\mu(\Theta)$ and $\Sigma_{TS}(\Theta)$ that correspond to the observed components of $\mathbf{x}_{TS}$. If for instance first $M$ out of $N$ components of $x$ are observed than $y(t) = (x_1(t), ..., x_M(t))$, and

$$\mu_y(\Theta) = (\tilde{\varphi}_M(t_1), ..., \tilde{\varphi}_M(t_n)) \tag{21}$$

where $\tilde{\varphi}_M(t) = (\tilde{\phi}_1(t), ..., \tilde{\phi}_M(t))$ and $\hat{\Sigma}(\Theta)$= is a $MN \times MN$ block matrix

$$\hat{\Sigma}(\Theta) = \left\{ \hat{\Sigma}(\Theta)^{(i,j)} \right\}_{i=1,...,N;j=1,...,N} \tag{22}$$

where

$$\hat{\Sigma}_{pq}^{(i,j)}(\Theta) = \Sigma_{pq}^{(i,j)}(\Theta) \quad p = 1, ..., M, q = 1, ..., M. \tag{23}$$

# 3 Calculation of the Fisher Information Matrix (FIM)

Suppose that a random variable $X$ has an $N$-variate normal distribution with mean $\mu(\Theta) = (\mu_1(\Theta), ..., \mu_N(\Theta))^T$ and covariance matrix $\Sigma(\Theta)$. We define the FIM for this variable to be $I(\Theta)$= $\{I(\Theta)_{k,l}\}$ [7]

$$I(\Theta)_{k,l} = E_\Theta \left[ \left( \frac{\partial}{\partial \theta_k} \log(\psi(X, \Theta)) \right) \left( \frac{\partial}{\partial \theta_l} \log(\psi(X, \Theta)) \right) \right], \tag{24}$$

where $\psi(.)$ is the density function of a multivariate normal distribution with mean $\mu(\Theta)$ and covariance $\Sigma(\Theta)$. As the random variable $X$ is normally distributed the elements $I(\Theta)_{k,l}$ can be also expressed explicitly as

$$I(\Theta)_{k,l} = \frac{\partial \mu}{\partial \theta_k}^T \Sigma(\theta) \frac{\partial \mu}{\partial \theta_l} + \frac{1}{2} trace(\Sigma^{-1} \frac{\partial \Sigma}{\partial \theta_k} \Sigma^{-1} \frac{\partial \Sigma}{\partial \theta_l}). \tag{25}$$

In this section we demonstrate how to calculate the FIM for the models (2- 4). We consider the model for time series data, as the case of time-point measurements is less general and can be directly extracted from formulas derived below. Previously, we have shown 15 that variable $\mathbf{x}_{TS}$ has a multivariate normal distribution and demonstrated how its mean $\mu(\Theta)$ and covariance matrix $\Sigma(\Theta)$ can be calculated. Formula (25) indicates that two more components $\frac{\partial \mu}{\partial \theta_k}$ and $\frac{\partial \Sigma(\Theta)}{\partial \theta_k}$ need to be known in order to be able to compute FIM. Below we show how these can be obtained.

Let $Y(t)$ be the concatenated vector of $\varphi(t)$ and upper diagonal of the symmetric matrix $V$

$$Y(t) = (\phi_1(t), ..., \phi_N(t), V_{1,1}(t), ..., V_{N,N}(t), ..., V_{1,2}(t), ..., V_{N-1,N}(t)) \tag{26}$$

and $\tilde{Y}(Y_0, \Theta, t)$ be the concatenation of $\tilde{\varphi}(\varphi_0, \Theta, t)$ and upper diagonal of $\tilde{V}(V_0, \Theta, t)$. Similarly denoting the concatenation of the right hand sides of equations (3) and (11) by $W$ we can write

$$\frac{d}{dt} Y(t) = W(Y(t), \Theta, t). \tag{27}$$

To determine the derivative $Z_k(t) = \frac{\tilde{Y}(t)}{\partial \theta_k}$ we use the fact that it satisfies the following equation (see Appendix)

$$\frac{d}{dt} Z_k(t) = J(\tilde{Y}(t), \Theta, t) Z_k(t) + K_k(t), \tag{28}$$

where $J(\tilde{Y}(t), \Theta, t)$ is the Jacobian $\frac{\partial}{\partial Y(t)} W(Y(t), \Theta, t)$ evaluated at the solution $\tilde{Y}(t)$ and $K_k(t)$ is the vector $\frac{\partial}{\partial \theta_k} W(Y(t), \Theta, t)$ also evaluated at $\tilde{Y}(t)$.
The solution of equation (28), $\tilde{Z}(t)$, provides us with $\frac{\partial \tilde{\phi}(t)}{\partial \theta_k}$ and therefore with $\frac{\partial \mu}{\partial \theta_k}$. Similarly $\tilde{Z}(t)$ contains diagonal elements of $\frac{\partial \Sigma}{\partial \theta_k}$.

Non-diagonal elements of $\frac{\partial \Sigma}{\partial \theta_k}$ can be computed from diagonal elements using the recursive relation

$$\frac{\partial}{\partial \theta_k} \Sigma^{(i,j+1)} = \frac{\partial}{\partial \theta_k} \left( \Sigma^{(i,j)} \Phi(t_j, t_{j+1})^T \right) = \frac{\partial}{\partial \theta_k} \left( \Sigma^{(i,j)} \right) \Phi(t_i, t_{i+1})^T + \Sigma^{(i,j)} \frac{\partial}{\partial \theta_k} \left( \Phi(t_i, t_{i+1})^T \right). \tag{29}$$

from elements $\Phi(t_j, t_{j+1})$, $\Sigma^{(i,i)}$, $\frac{\partial}{\partial \theta_k} \Sigma^{(i,i)}$ that are given by equations (9), (16) and (28) respectively. To simplify notation denote $\Xi_k(s, t) = \frac{\partial \Phi(s,t)}{\partial \theta_k}$. As $\Phi(s, t)$ is a solution of an ODE we use similar techniques as in equations (28) and write $\Xi_k(s, t)$ as a solution of the

differential equation

$$\frac{d\Xi_k}{dt}(s,t) = A(\tilde{\varphi}, \Theta, t)\Xi(s,t) + M_k(t), \tag{30}$$

where

$$M_k(t) = \frac{\partial}{\partial\theta_k}(A(\tilde{\varphi},\Theta,t)\Phi(s,t)) = \left(\frac{\partial}{\partial\theta_k}A(\tilde{\varphi},\Theta,t) + (\frac{\partial}{\partial\varphi}A(\varphi,\Theta,t)_{\varphi=\tilde{\varphi}})\frac{\partial\tilde{\varphi}}{\partial\theta_k}\right)\Phi(s,t), \tag{31}$$

and $\Xi(s,s) = 0$ for all $s$. To summarise, for the experimental data distribution (15) the FIM (25) can be computed using equations (28 - 31):

- the parameter derivative of the mean, $\frac{\partial\mu(\Theta)}{\partial\theta_k}$, can be extracted from a solution of (28)

- diagonal elements of the parameter derivatives of the variance, $\frac{\partial\Sigma_{TS}(\Theta)}{\partial\theta_k}$, can be extracted from a solution of (28)

- non-diagonal elements of parameter derivatives of the variance, $\frac{\partial\Sigma_{TS}(\Theta)}{\partial\theta_k}$, are given by formula (29), which involves (30) and (31).

## 3.1   Summary of the numerical computation of the FIM

Below we summarise in more details how the FIM can be calculated numerically. We start with the case of time series data as it is most general and the remaining two can be derived from it.

**Time series measurmens**

1 Read input: Stoichiometry matrix $S$, reaction rates vector $F$, initial conditions $x_0$, $V_0$

2 Construct equations for $\varphi$ (eq. (3)) and $V$ (eq. (11)) and for $Y$ (eq. (27))

3 Calculate symbolically the Jacobians $A$ (eq. (6)), $J$ (eq. (28)) and vectors $K_k$ (eq. (28)) , $M_k$ (eq. 30) $(k = 1, ..., L)$

4 Solve equations for $\varphi$ and $V$ (eq. (27))

5 Compute fundamental matrices $\Phi(t_i, t_{i+1})$ $i = 1, ..., N-1$ (eq. (9))

6 Construct covariance matrix $\Sigma_{TS}$ from $\tilde{V}(t_i)$ and $\Phi(t_i, t_{i+1})$ $(i = 1, ..., n)$ according to eq. (16)

7 Compute $\frac{\partial \tilde{Y}}{\partial \theta_k}$ (solve eq. (28)) and extract $\frac{\partial \mu}{\partial \theta_k}$ $\frac{\partial \tilde{V}}{\partial \theta_k}$ according to eq. (26)

8 Compute $\frac{\partial}{\partial \theta_k} \Phi(t_i, t_{i+1})$ (eq. (30))

9 Compute $\frac{\partial}{\partial \theta_k} \Sigma^{(i,j)}$ for $j = \geq i+1, ..., n$ and $i = 1, ..., n$ (eq. (29))

10 Construct $\frac{\partial}{\partial \theta_k} \Sigma_{TS}$ from objects computed in steps 7 and 9

11 From $\frac{\partial}{\partial \theta_k} \varphi$, $\Sigma_{TS}$, $\frac{\partial}{\partial \theta_k} \Sigma_{TS}$ extract those elements corresponding to observed component
according to relations (21), (22) and (23)

12 Compute FIM from elements obtained in the previous steps according to eq.
(25)

**Time-point measurmens**

In order to compute the FIM for time-point measurements the covariance matrix, $\Sigma_{TS}$, should be replaced by $\Sigma_{TP}$. Additionally non-diagonal blocks of covariance matrix, $\Sigma_{TP}$, are equal to 0, therefore steps 5, 8 and 9 are omitted and in step 3 vectors $M_j$ need not be computed.

**Deterministic model data**

For the deterministic model the covariance matrix does not depend on parameters, therefore the formula for the elements of FIM simplifies to

$$I(\Theta)_{k,l} = \frac{\partial \mu}{\partial \theta_k}^T \Sigma_{DT}(\Theta) \frac{\partial \mu}{\partial \theta_l},$$ \hfill (32)

and it requires only calculation of derivatives $\frac{\partial \tilde{z}}{\partial \theta_k}$ for $k = 1, \ldots, L$.

# 4 Examples

In this section we present details pertaining to the examples of models of single gene expression and the p53 system.

## 4.1 A model of a single gene expression

The Table 1 contains parameter values used for numerical experiments presented in the main paper.

| Param. | Set 1 | Set 2 | Set 3 | Set 4 |
|--------|-------|-------|-------|-------|
| $k_r$ | 100 | 100 | 20 | 20 |
| $k_p$ | 2 | 2 | 10 | 10 |
| $\gamma_r$ | 1.2 | 0.7 | 1.2 | 0.7 |
| $\gamma_p$ | 0.7 | 1.2 | 0.7 | 1.2 |

Table 1: Four parameter sets used in analysis of the single gene expression model. Sets 1 and 3 correspond to slow protein degradation rate $\gamma_p$ and high and low transcription / translation ratio, respectively. On the other hand Sets 2 and 4 describe fast protein degradation rate and high and low transcription / translation ratio, respectively. All rates are per hour.

### 4.1.1 Differences in sensitivity and robustness analysis between time-series, time-points and deterministic versions of the model

Considering sensitivity and robustness analysis, there are three main differences between stochastic and deterministic systems. Firstly deterministic models completely neglect variability in the abundances of molecular species. This variability is a function of the kinetic parameters and is therefore also sensitive to them. Secondly, the deterministic model does

| Type | # ident. param. | $CR(\log(k_r))$ | $CR(\log(k_p)))$ | $CR(\log(\gamma_r))$ | $CR(\log(\gamma_p))$ | det( FIM ) |
|------|-----------------|-----------------|------------------|----------------------|----------------------|------------|
| $TS$ | 4 | 0.0017 | 0.0016 | 0.0017 | 0.0017 | $4 \cdot 10^3$ |
| $TP$ | 2 | 0.0002 | 0.0002 | 0.0002 | 0.0002 | 0 |
| $DT$ | 1 | $3 \cdot 10^{-3}$ | $3 \cdot 10^{-3}$ | $3 \cdot 10^{-3}$ | $3 \cdot 10^{-3}$ | 0 |

Table 2: Identifiability analysis for stationary state data. The table presents the number of non-zero eigenvalues (# ident. param.), Cramer-Rao bounds (CR), determinants of FIM (det(FIM)) for different data types (time series (TS), time point measurements (TP), deterministic model (DT)). The number of non-zero eigenvalues equals to the number of (in principle) identifiable linear combinations of parameters and therefore describes the number of parameters that can be estimated given that others are known. Quantities were calculated for parameter set 3 (see Table 1) and we have set the sampling frequency to $\Delta = 0.3h$, and the number of measurements to $n = 50$. The system was supposed to be in the stationary state. We have assumed that a parameter is identifiable if an eigenvalue of FIM is not lower than $10^{-4}$ to take account of numerical inaccuracies, and therefore "# ident. param." is calculated as the number of eigenvalues that are greater or equal to 0.1% of the largest eigenvalue. For the same reason the determinant was calculated as the product of eigenvalues that satisfies this condition. As not all parameters were identifiable in all versions of the model we calculated CR for individual estimates (assuming all other parameters to be known). For the deterministic model we have set variance of measurement error $\sigma_\epsilon^2 = 100$ and no measurement error for TS and TP therefore CR-bounds between stochastic and deterministic models cannot be compared.

not include correlations between molecular species. Thirdly, temporal correlations are also neglected.

In the main paper we argued that these three factors can have a significant impact on how stochastic and deterministic systems respond to perturbations in parameters. Here we provide further explanation using the model of single gene expression. The formulae for mean,

| Type | # ident. param. | $CR(\log(k_r))$ | $CR(\log(k_p))$ | $CR(log(\gamma_r))$ | $CR(\log(\gamma_p))$ | det(FIM) |
|------|-----------------|-----------------|-----------------|---------------------|----------------------|----------|
| $TS$ | 4 | 0.0413 | 0.0112 | 0.0098 | 0.0072 | $6.96 \cdot 10^4$ |
| $TP$ | 4 | 0.0185 | 0.0036 | 0.0036 | 0.0020 | $6.94 \cdot 10^3$ |
| $DT$ | 3 | $0.47 \cdot 10^{-4}$ | $0.04 \cdot 10^{-4}$ | $0.07 \cdot 10^{-4}$ | $0.02 \cdot 10^{-4}$ | 0 |

Table 3: Identifiability analysis for perturbation experiment. Identical analysis as in Table 2 but with an initial mean increased 5 fold and the initial variance 25 fold.

variances and covariance for this model are

$$\langle r \rangle = \frac{k_r}{\gamma_r} \tag{33}$$

$$\langle p \rangle = \frac{k_r k_p}{\gamma_r \gamma_p} \tag{34}$$

$$\langle \delta r^2 \rangle = \frac{k_r}{\gamma_r} \tag{35}$$

$$\langle \delta p^2 \rangle = \langle p \rangle (1 + \frac{k_p}{\gamma_r + \gamma_p}) \tag{36}$$

$$\langle \delta r \delta p \rangle = \frac{k_p k_r}{(\gamma_r + \gamma_p) \, \gamma_r}. \tag{37}$$

In order to understand the effect of incorporating variability into the sensitivity analysis we are considering changes in parameters, e.g. $k_p, \gamma_p$, by a factor $\delta$ $(k_p, \gamma_p) \rightarrow (k_p + \delta k_p, \gamma_p + \delta \gamma_p)$. Means of RNA and protein concentrations are not affected by this perturbation, whereas protein variance is.

To understand the effect of correlation between RNA and protein levels we note that formulae (33 - 37) demonstrate that at constant mean, correlation increases with $\gamma_p$ and compensating decrease in $k_p$. Figure 1 presents neutral spaces for all parameter pairs for different values of correlation $\rho_{rp} = \frac{\langle \delta r \delta p \rangle}{\sqrt{\langle \delta r^2 \rangle \langle \delta p^2 \rangle}}$. Differences between deterministic and stochastic model increase with correlation.

We also perform similar analyses for different levels of temporal correlation between observations by varying the sampling frequency $\Delta$. Figure 2 presents neutral spaces for all parameter pairs for three different sampling frequencies. The differences between stochastic and deterministic models decrease with $\Delta$ as the samples are less correlated for high $\Delta$, and therefore the factor that distinguishes two models becomes less significant.

11

## 4.2 P53 system

In the study of the 53 system we have used parameter estimates presented in Table 4. These parameters has been obtained by appropriate scaling of the parameters given in [8]. For all numerical experiments for p53 model we assumed sampling frequency $\Delta = 1h$ and number of measurements $n = 30$.

| Param. | Value |
|--------|-------|
| $\beta_x$ | 90 |
| $\alpha_x$ | 0.002 |
| $\alpha_k$ | 1.7 |
| $k$ | 0.01 |
| $\beta_y$ | 1.1 |
| $\alpha_0$ | 0.8 |
| $\alpha_y$ | 0.8 |

Table 4: Parameters of p53 system.

### 4.2.1 Sloppiness analysis

Here we compare neutral spaces for all pairs of parameters of the P53 model for tree data types (TS, TP, DT). We use parameter values presented in Table 4 and logarithmic parametrisation. Results are presented in Figues 3, 4 and 5.

## 4.3 Dependance of analysis on parameter values and qualitative model behaviour

Our method allows to study model sensitivity given the parameter values. Here we show that results depend on parameter values just as the dynamical behaviour of the system does; i.e. the sloppiness of a system also depends on the parameters and is not a fixed property of a mathematical model. Figure 8 demonstrates that p53 undergoes a Hopf bifurcation as parameter $\alpha_y$ is varied from 0.8 to 2, while all other parameters remain unchanged. Parameters thus determine the qualitative dynamical behaviour and therefore varying parameters

influences the structure of the FIM (compare Figures 3 and 9). Change in the FIM in turn has consequence for the magnitude of eigenvalues (Figure 8), sensitivity coefficients (compare Figures 4 in **MP** and 11) and informational content of TS and TP data (Figures 7 and 12).

# 5 Appendix

## 5.1 Calculating derivatives of a solution of ODE

In order to derive equations (28) and (30) we differentiated the solution of an ODE with respect to a parameter. Here we provide more details about this procedure. Suppose that the differential equation being considered is

$$\dot{x} = \mathsf{F}(x, \Theta, t),$$

where $x \in R^N$ and the set of parameters are collected together into a parameter vector $\Theta \in R^L$. Suppose that $\tilde{x}(\Theta, x_0, t)$ is the solution of interest with an initial condition $x_0 = \tilde{x}(\Theta, x_0, 0)$. For $Y(t) = \frac{\partial \tilde{x}(\Theta, x_0, t)}{\partial \theta_k}$ it can be shown [9] that

$$\dot{Y} = J(t)Y(t) + K_k(t) \tag{38}$$

where $J(t)$ is the Jacobian of $F$ with respect to $x$ evaluated at $\tilde{x}$, $K_k(t)$ is the n-dimensional vector $\frac{\partial F}{\partial \theta_k}$ and $Y(0) = 0$.

## 5.2 Logarithmic parametrisation

In the analysis of examples presented in our study we used a logarithmic parametrisation. Below we provide a rationale for this and explain that the FIM for logarithmic parametrisation can be directly obtained from derivatives calculated to obtain the FIM for original parameters.

In biochemical systems, the values of two parameters may differ by orders of magnitudes. Therefore, it is usually not appropriate to consider the absolute changes in the parameters $\theta_k$, but instead to consider the relative changes. A good way to do this is to introduce new parameters $\eta_k = \log(\theta_k)$, because absolute changes in $\eta_k$ correspond to relative changes in $\theta_k$. Then, for small changes $\delta\theta_k$ to the parameters $\theta_k$, the changes $\eta_k$ are scaled and non-dimensional. Analyses in terms of absolute and relative changes are closely related and do not require additional computational cost. For any differentiable function $f(\theta)$

$$\frac{\partial f}{\partial \log(\theta)} = \frac{\partial f}{\partial \eta} = \theta \frac{\partial f}{\partial \theta}. \tag{39}$$

13

and therefore any element of FIM for $\eta$ can be easily converted into that for $\theta$ or *vice versa*

$$I(\eta)_{k,l} = \frac{\partial \mu}{\partial \eta_k}^T \Sigma(\eta) \frac{\partial \mu}{\partial \eta_l} + \frac{1}{2} trace(\Sigma^{-1} \frac{\partial \Sigma}{\partial \eta_k} \Sigma^{-1} \frac{\partial \Sigma}{\partial \eta_l})$$

$$= \theta_k \theta_l \frac{\partial \mu}{\partial \theta_k}^T \Sigma(\theta) \frac{\partial \mu}{\partial \theta_l} + \theta_k \theta_l \frac{1}{2} trace(\Sigma^{-1} \frac{\partial \Sigma}{\partial \theta_k} \Sigma^{-1} \frac{\partial \Sigma}{\partial \theta_l})$$

## 5.3 The FIM as a measure of system's sensitivity

Here we provide an alternative explanation why the FIM provides a measure of sensitivity for a stochastic system. For notational simplicity we assume that the studied system depends on a single parameter $\theta$ as generalisation for multidimensional parameter is straightforward. We start with definitions of classical sensitivity coefficients.

### 5.3.1 Classical sensitivity coefficient

The classical sensitivity coefficient for observable $Q$ and parameter $\theta$ is defined as [10]

$$S = \frac{\partial Q}{\partial \theta}.$$

Often sensitivity of relative changes $\frac{\Delta Q}{Q}$ needs to be considered. Given that $\Delta \log(Q) \approx \frac{\Delta Q}{Q}$ the formula for sensitivity of relative changes takes the form

$$S = \frac{\partial \log(Q)}{\partial \theta}.$$

### 5.3.2 The FIM as a sensitivity measure for a stochastic system

The behaviour of a stochastic system is not defined by an observable $Q$ that can be measured experimentally in a reproducible way. It is instead defined by a distribution form which the measurements are taken. Suppose we want to construct a measure of a sensitivity of a distribution of a random variable $X$ with density $\psi$. Assume we want to examine relative changes of the distribution $\psi$ to changes in $\theta$. This can be written as

$$\frac{\psi(X, \theta + \partial \theta) - \psi(X, \theta)}{\psi(X, \theta)} \simeq \log(\frac{\psi(X, \theta + \partial \theta)}{\psi(X, \theta)}).$$

Averaging over all possible observations we get

$$\int_X \log(\frac{\psi(X, \theta + \partial \theta)}{\psi(X, \theta)}) \psi(X, \theta) dX.$$

14

The above is the negative Kullback-Leibler divergence between distributions $\psi(X, \theta + \partial\theta)$ and $\psi(X, \theta)$. In order to study changes in $\psi$ resulting from "small" changes in $\theta$ we divide the above equation by $\partial\theta$ and take the limit $\partial\theta \to 0$ and get

$$\int_X \frac{\partial \log(\psi(X, \theta))}{\partial\theta} \psi(X, \theta) dX.$$

The above quantity is the average of the score function and it is basic fact of mathematical statistics that it equals to zero [7]. This observation suggests that it is better to study the squared differences

$$\int_X (\log(\psi(X, \theta) - \log(\psi(X, \theta + \partial\theta))^2 \psi(X, \theta) dX,$$

that lead to

$$\int_X \frac{(\log(\psi(X, \theta) - \log(\psi(X, \theta + \partial\theta))^2}{(\partial\theta)^2} \psi(X, \theta) dx \xrightarrow[\partial\theta \to 0]{} \int_X \left( \frac{\partial}{\partial\theta} \log\left( \frac{(\psi(X, \theta + \partial\theta))}{\psi(X, \theta)} \right) \right)^2 \psi(X, \theta) dx,$$

which is precisely the definition of the FIM.

The above derivation suggests that the FIM is a good measure of sensitivity of a probability distribution and that there is a close link between Kulback-Leibler divergence and the FIM. The KL divergence measures the average relative difference between two distributions whereas FIM measures squared relative difference between a distribution and the same distribution with a perturbed parameter relative to the infinitesimal size of the squared perturbation.

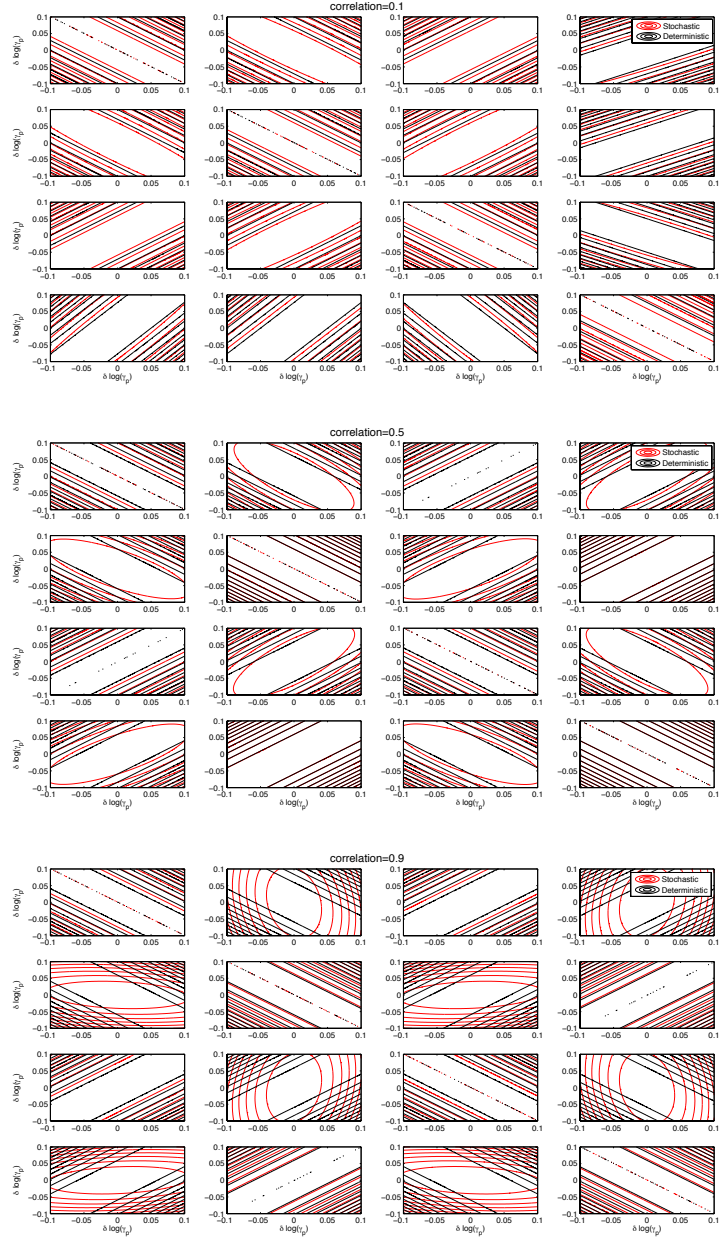

Figure 1: FIM for a single measurement from stationary distribution of the model of single gene expression. For $\rho_{rp} = 0.1$ (top), $\rho_{rp} = 0.5$ (middle), $\rho_{rp} = 0.9$ (bottom). Correlation 0.5 was obtained using parameter set 3 from Table 1. Correlation was varied by equal-scaling of parameters $k_p, \gamma_p$.

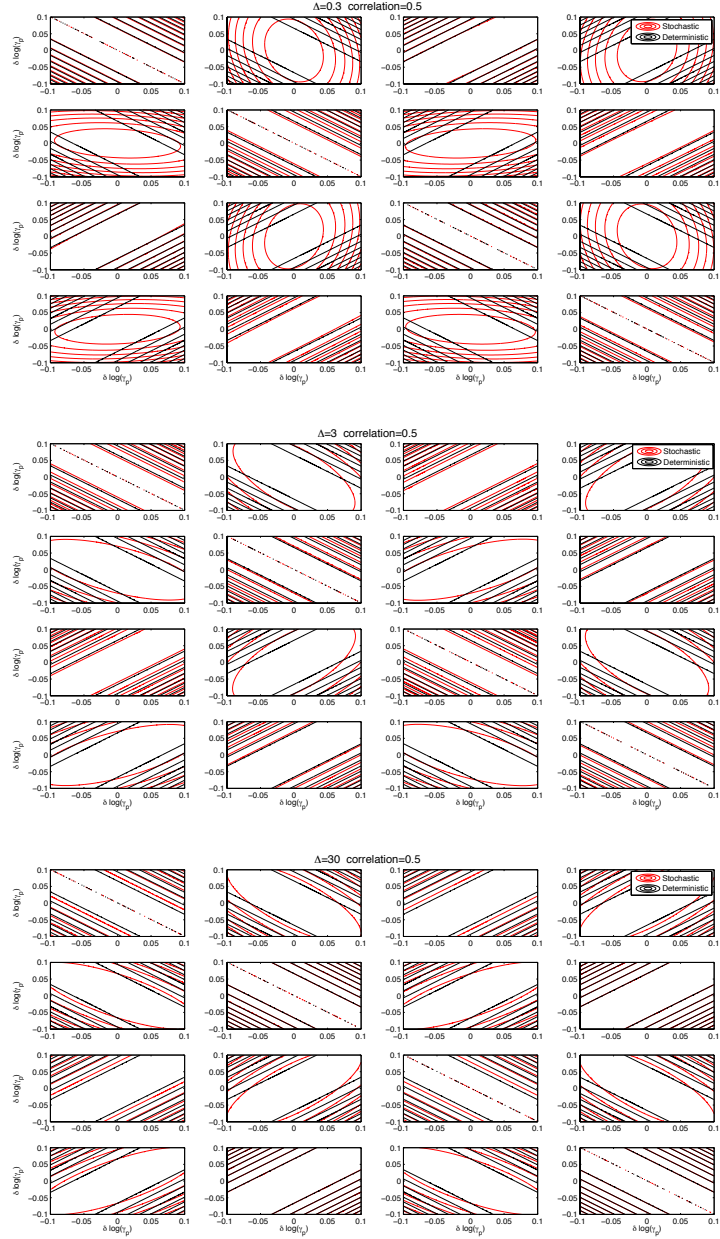

Figure 2: FIM for a 20 times-series type measurements from the stationary distribution of the model of single gene expression for three different sampling frequencies: $\Delta = 0.3$ (top) $\Delta = 3$ (middle) $\Delta = 30$ (bottom). Parameter set 3 from Table 1 was used.

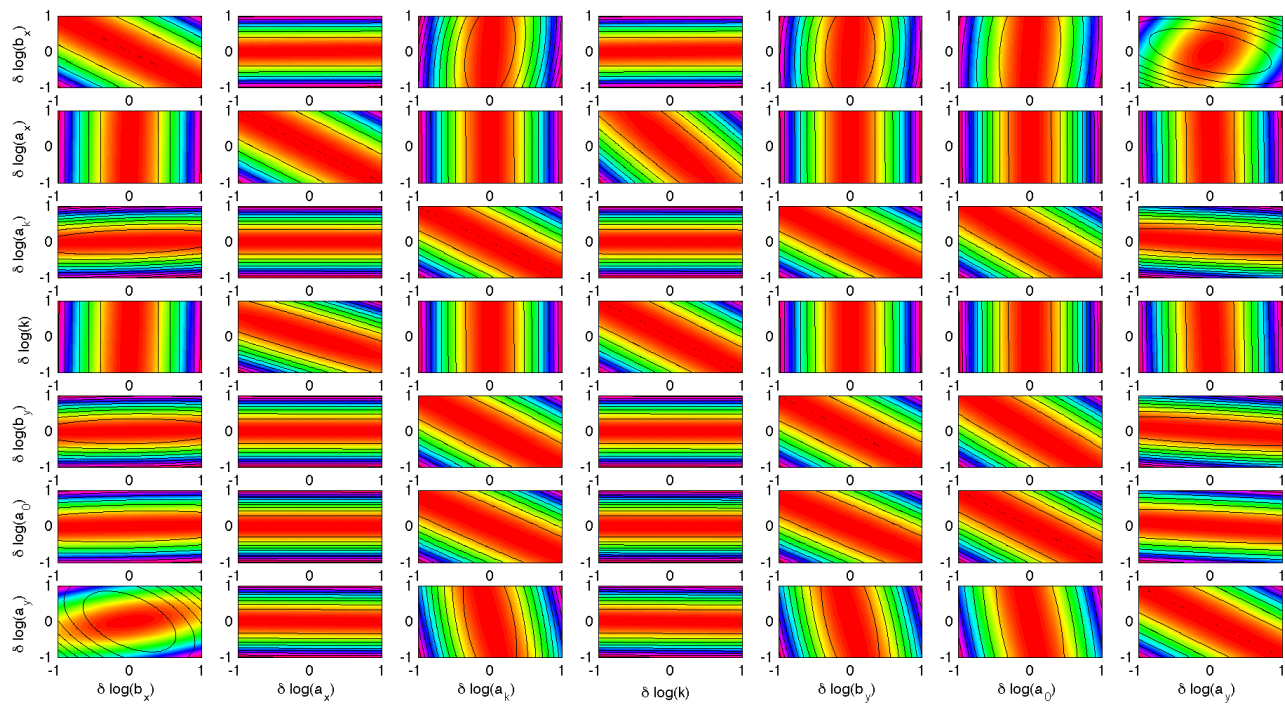

Figure 3: Neutral space for time-series (heatmap) and deterministic (contour plot) versions of the p53 model. The FIM was calculated for the logarithms of parameters in Table 4.

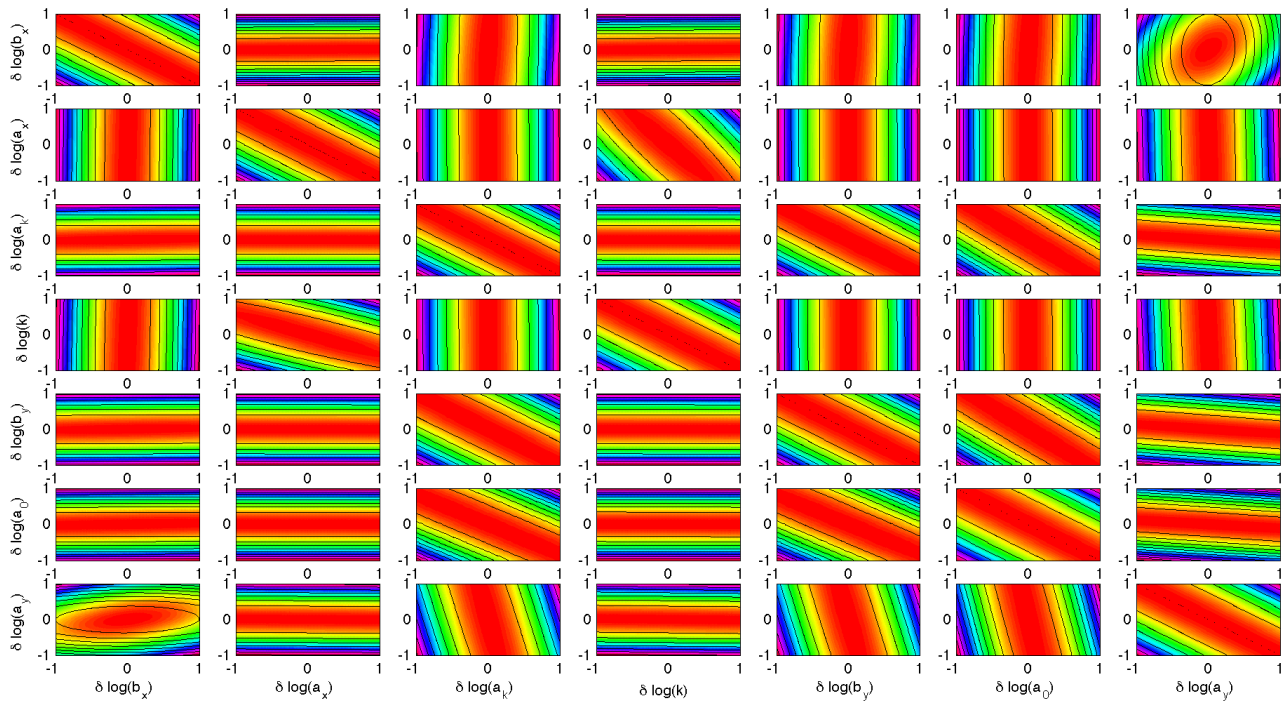

Figure 4: Neutral space for time-series (heatmap) and time-point (contour plot) versions of the p53 model. The FIM was calculated for logarithms of parameters in Table 4.

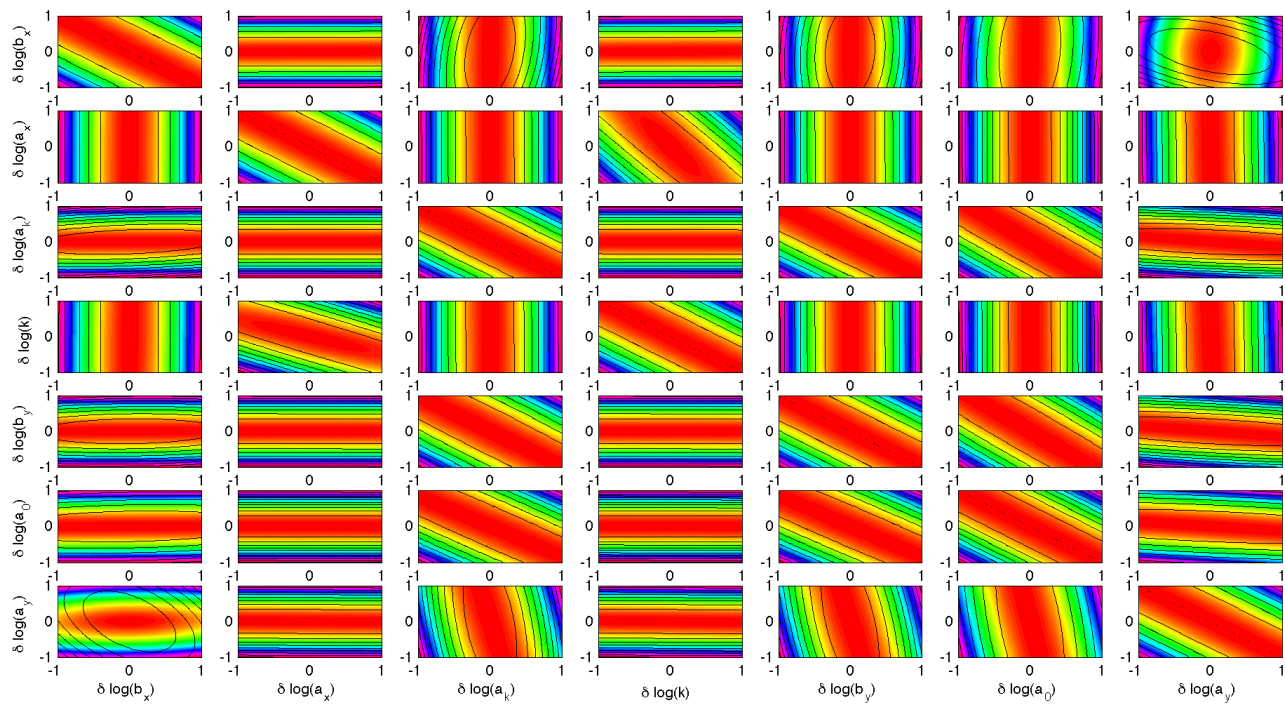

Figure 5: Neutral space for time-points (heatmap) and deterministic (contour plot) versions of the p53 model. The FIM was calculated for logarithms of parameters in Table 4.

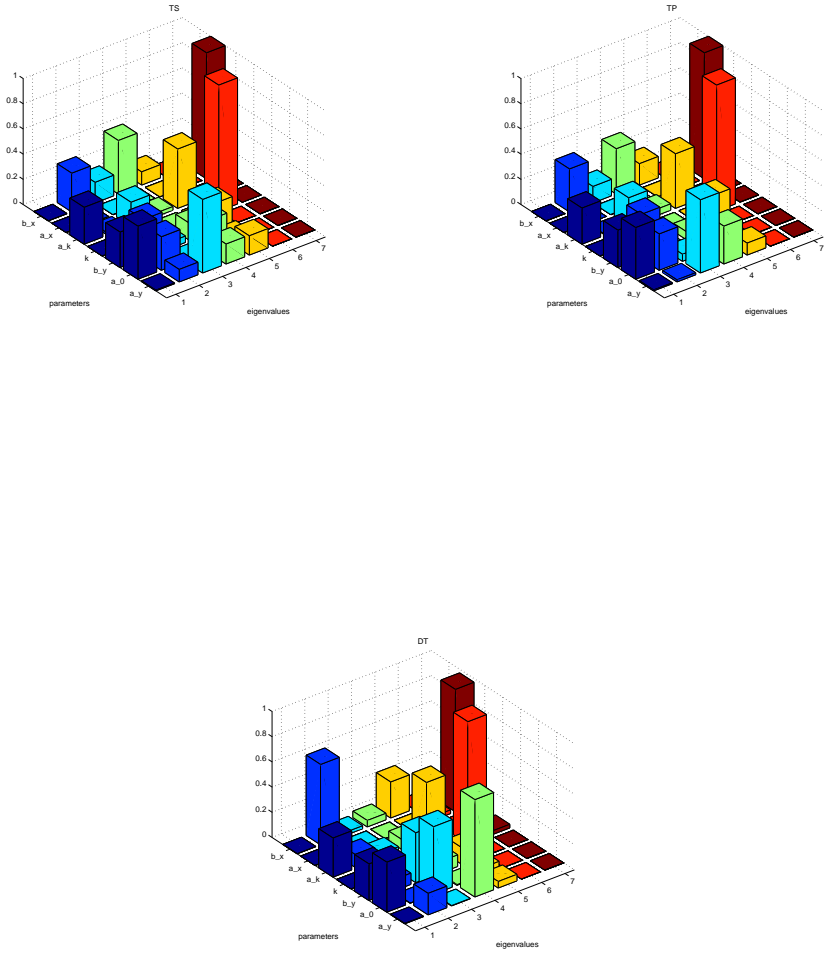

Figure 6: Sensitivity matrices $C_{ij}^2$ for p53 model for three data types (TS, TP, DT) calculated using parameters presented in Table 4.

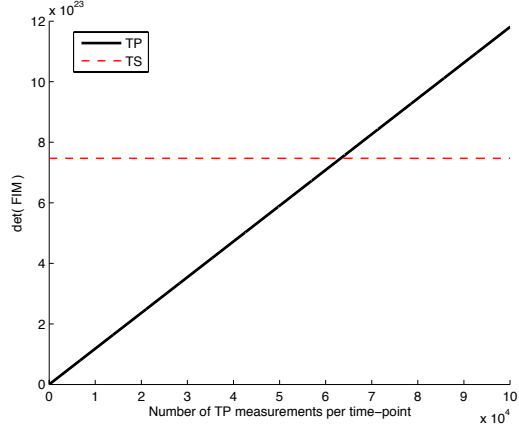

Figure 7: Comparison of informational content of TP and TS samples for p53 model. Determinant of the FIM for TP data is plotted against number of measurements per time point (black line). Due to independence of measurements we observe the linear increase. For comparison determinant of the FIM for a single TS sample is also depicted (red dashed line). Intersection of the two lines indicates how many TP measurements are necessary to obtain the same amount of information in a single trajectory (TS data). In this case around $6.5 \cdot 10^4$ measurements are needed.

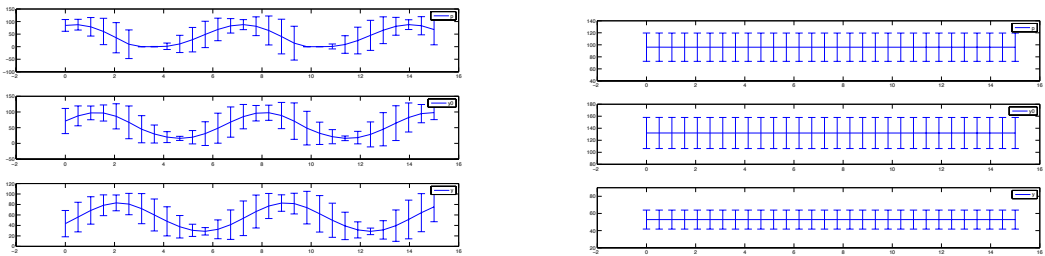

Figure 8: Trajectories of the p53 system plotted together with the standard deviations bars. In the left panel parameters from Table 4 were used and in the right panel the same parameters except $\alpha_y = 2$. The system undergoes a Hopf bifurcation and moves from oscillatory behaviour to dynamics with a stable stationary state.

23

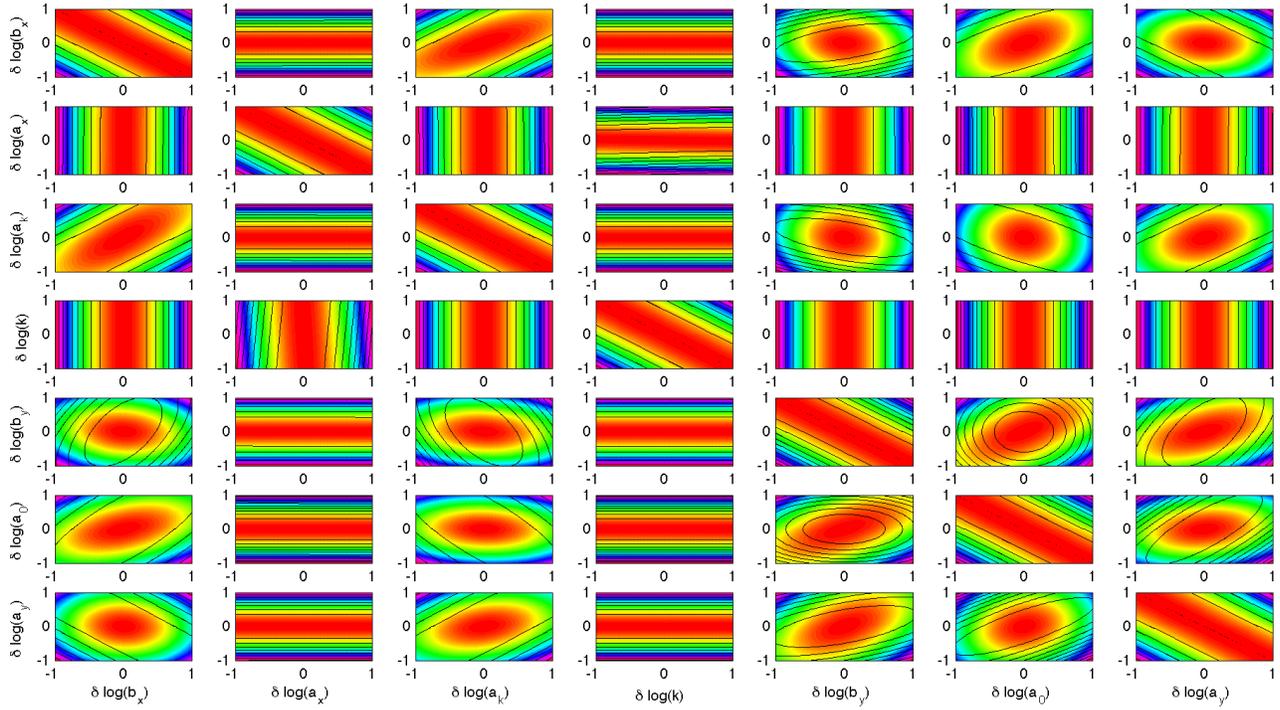

Figure 9: Neutral spaces for time-points (heatmap) and deterministic (contour plot) versions of the p53 model. FIM was calculated for logarithmss of parameters in Table 4 except $\alpha_y = 2$. Differences compared with Figure 5 demonstrate the dependance of neutral spaces on parameter values and the qualitative dynamics of the system.

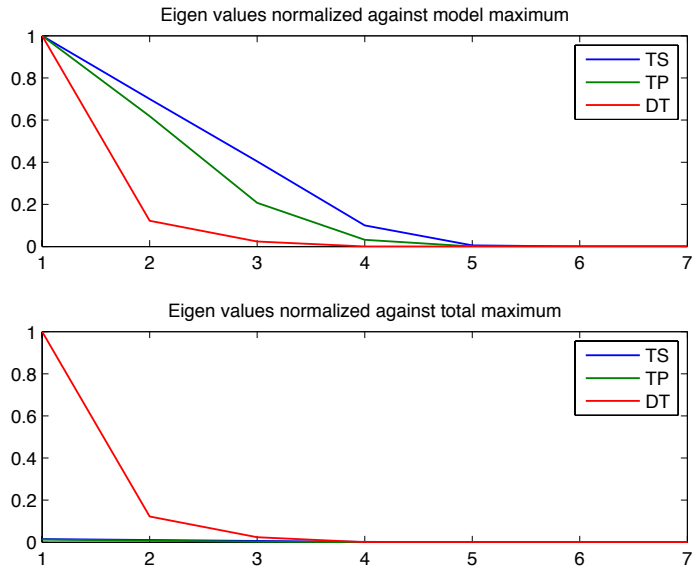

Figure 10: Eigenvalues of FIM for p53 model for three data types: time series (blue), time-points (green) and deterministic model (red). Eigenvalues were normalised against maximal eigenvalue for each data type (top) and against maximal eigenvalue among all three types (bottom). FIM was calculated for logs of parameters from Table 4 except $\alpha_y = 2$. Figure demonstrates that the behaviour of the eigenvalues depends on parameter values (compare with Figure 3 in the **MP**).

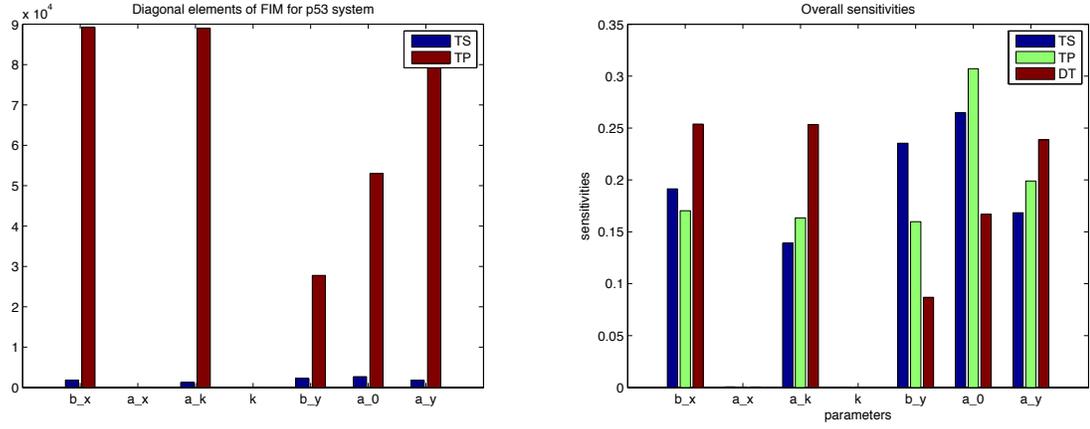

Figure 11: **Left:** Diagonal elements of FIM for TS and TP versions of p53 model. **Right:** Sensitivity coefficients $\mathcal{T}_i$ for TS, TP, DT version of p53 model. FIMs were calculated for parameters presented in Table 4 except $\alpha_y = 2$.

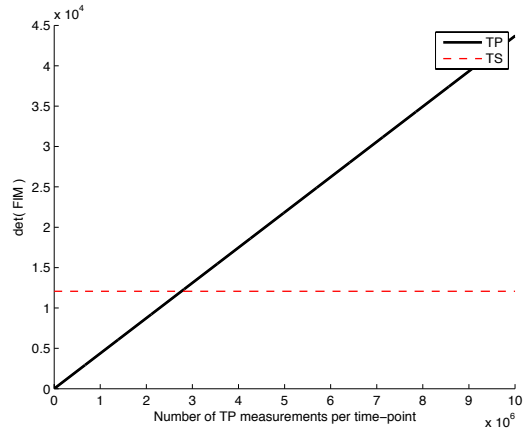

Figure 12: Comparison of informational content of TP and TS samples for p53 model similarly as Figure 7 but with $\alpha_y = 2$ instead of $\alpha_y = 0.8$.



\end{document}